\documentclass[sigconf]{acmart}

\AtBeginDocument{%
  \providecommand\BibTeX{{%
    \normalfont B\kern-0.5em{\scshape i\kern-0.25em b}\kern-0.8em\TeX}}}
\acmConference{}{}{}

\newcommand\cancel[1]{}

\newtheorem{openq}{Open Question}

\usepackage{balance}

\usepackage{color}
\usepackage[T1]{fontenc}
\usepackage{lmodern}
\definecolor{lightgray}{gray}{0.75}

 \renewcommand\footnotetextcopyrightpermission[1]{} 
 \setcopyright{none}
 \acmDOI{}
 \acmISBN{}
 \acmPrice{}
\usepackage[noframe]{showframe}
\usepackage{framed}
\usepackage{lipsum}
 {\endMakeFramed}
\definecolor{shadecolor}{gray}{0.85}

\usepackage{ifthen}
 
\newboolean{SHORT}
\setboolean{SHORT}{false}

\begin{document}

\title[Cryptocurrency Networking]{Survey on Cryptocurrency Networking:\\
Context, State-of-the-Art, Challenges}


\author{Maya Dotan$^1$ \quad
 Yvonne-Anne Pignolet$^2$ \quad
Stefan Schmid$^3$  \quad
Saar Tochner$^1$ \quad 
Aviv Zohar$^1$}
\affiliation{$^1$ Hebrew University, Israel \quad $^2$ DFINITY, Switzerland \\ $^3$ Faculty of Computer Science, University of Vienna, Austria 
}

\renewcommand{\shortauthors}{Dotan et al.}

\begin{abstract}
Cryptocurrencies such as Bitcoin are realized using distributed 
systems and hence critically rely on the performance
and security of the interconnecting network.
The requirements on these networks and their usage, 
however can differ significantly from traditional communication networks,
with implications on all layers of the protocol stack.
This paper is motivated by these differences, and in particular
by the observation that many fundamental design aspects of these
networks are not well-understood today. 
In order to support the networking community to contribute
to this emerging application domain, 
we present a structured overview of the field,
from topology and neighbor discovery
to block and transaction propagation. 
In particular, we provide the context, 
highlighting differences and commonalities
with traditional networks,
review the state-of-the-art, and identify
open research challenges. 
Our paper can hence also be seen as a call-to-arms to improve 
the foundation on top of which cryptocurrencies are built. 
\end{abstract}

\maketitle
\section{Introduction}\label{sec:intro}
Cryptocurrencies provide means to exchange digital assets relying on strong cryptography and, 
in contrast to centralized digital currencies and central banking systems, offer \emph{decentralized} control. 
Cryptocurrencies are thus typically realized
as distributed systems.
Accordingly, much existing research in the blockchain field focused on the underlying cryptographic primitives and on improved distributed blockchain protocols, e.g., consensus.
 
The network required to connect the distributed system, however,
has received relatively little attention. However, there is increasing evidence that the
network can become the bottleneck
and root-cause for some of the most pressing challenges blockchains face today. 
For example, the propagation of transactions and blocks (or other control messages for
the execution of consensus algorithms), require unicast and multicast communication services.  Blockchain miners which write transactions to the blockchain, are connected through
dedicated miner P2P networks~\cite{fibre}, in addition to the public blockchain P2P network.
Studies show that the cryptocurrency network layer
is critical for scalability~\cite{decker2013information,bloxroute}, 
security~\cite{gervais2016security} and privacy~\cite{gervais2014privacy} of a blockchain,
and that an efficient network layer 
enables higher transaction throughput and stronger resilience against malicious actors~\cite{gervais2016security}. 
Networking issues are also not limited to overlays on the network layer. Besides node discovery and data routing, the provided network functionality
for example includes the encoding and transmission of data, and error correction, as well as measurements of the network performance. 

Interestingly, cryptocurrency networks and their usage, can differ significantly from traditional communication networks.
For example, cryptocurrency networks may come with different requirements (e.g., related to anonymity), may need to serve different traffic mixes (e.g., more frequent broadcast of transactions and states of the blockchain), may need different routing mechanisms (e.g., source routing), or may be more dynamic (e.g., channels and fees in PCNs). Obviously, the fact that no protocol participant can be trusted (not only users issuing transactions but also miners 
and thus information providers may act maliciously), often requires a different design, and the incentives of all participants must be considered.

This paper is a call-to-arms to the networking community to identify the unique requirements of cryptocurrency networks and address the open issues.


\noindent \textbf{Our Contributions.}
This paper aims to provide a fast introduction
to cryptocurrency network issues with 
a focus on open research challenges.
To this end, we provide a structured overview of 
cryptocurrency networks, introducing 
the different aspects and their context, 
high-lighting differences and commonalities
with traditional networks,
and reviewing the state-of-the-art.
At the end of each section, we identify
and discuss research questions. 

This paper hence specifically targets 
junior and senior researchers with different backgrounds
(e.g., in networking, algorithms, or game theory) who would like
to get an overview of the state-of-the-art and 
start working in this area.  
It can also serve experts and decision-makers in the networking industry
as well as interested laymen.

\noindent \textbf{Related Work.}
This paper surveys
cryptocurrency network issues
with a focus on research challenges.
The most closely related papers in this
area are surveys and systematizations
of knowledge efforts 
on cryptocurrencies and blockchains in general~\cite{bonneau2015sok}.
Gudgeon et al.~\cite{gudgeon2019sok}
provide a comprehensive survey of
offchain networks, and  
Neudecker et al.~\cite{neudecker2018network} survey
the network layer of permissionless blockchains,
with a focus on attacks and the design space (but less
on research questions, e.g., revolving around incentives
or mining).
Katkuri presents a survey of data transfer and storage techniques in prevalent cryptocurrencies and suggests improvements~\cite{katkuri2018survey}.
The focus is on aspects related to the broadcast networks
underlying Ethereum, Nano and IOTA.
Gervais et al.~\cite{gervais2016security} give a thorough security analysis of
proof-of-work blockchain systems; their focus is
on the consensus layer (i.e., block generation) whereas the
network layer is abstracted. Troncoso et al.~\cite{troncoso2017systematizing} 
show a
broader perspective covering numerous systems apart from
Bitcoin and Tor but also abstract from the network layer. A
recent paper by Delgado-Seguara et al.~\cite{delgado2018cryptocurrency} 
explores the characteristics
of the peer-to-peer network established by Bitcoin,
but abstracts from the design space of the network layer.
Delgado et al. \cite{delgado2018cryptocurrency} 
provide an in-depth study of the Bitcoin P2P network. 

\begin{figure}[t]
	\centering
	\hspace{-.4cm}
	\includegraphics[width=1.0\columnwidth]{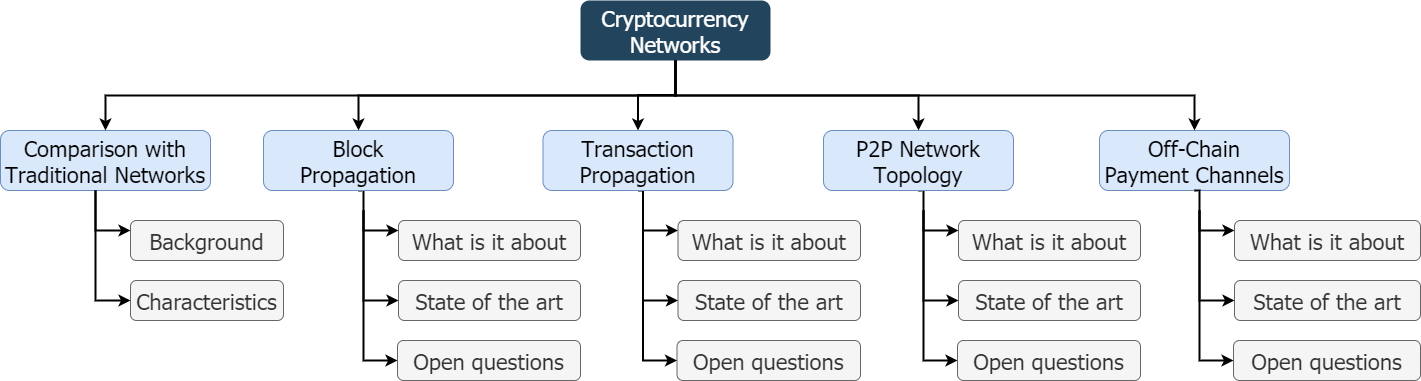}
	\vspace{-.2cm}
	\caption{Paper Organization}
	\label{figure::overview}
	\vspace{-.5cm}
\end{figure}

\noindent \textbf{Organization.}
The remainder of this paper is organized
as follows (see Figure~\ref{figure::overview}).
Section~\ref{sec:networksCryptocurency} provides an overview of some basic aspects of cryptocurrency networks. These include incentives, topology, communication pattern and security. Next, we discuss core aspects, namely block propagation, transaction propagation, P2P network topologies,
and off-chain networks in Sections~\ref{sec:blockpropagation}--\ref{sec:payment_channels}. In each of these sections, we give some background information, present the state of the art followed by open research questions that the writers of this paper find interesting. Finally, in Section~\ref{sec:conclusions} we conclude this SOK.


\section{Cryptocurrency Networks in Perspective}\label{sec:networksCryptocurency}

\subsection{Background} 
Before delving into the details, we provide some background
and introduce preliminaries.

\textbf{Cryptocurrencies} permit mutually distrusting parties to engage in financial
operations securely.
They guarantee that a transaction issued by Alice to send money to Bob reaches its destination at most once and only if Alice's balance is sufficiently high.
Analogously, operations with multiple senders and receivers are typically supported. 
The guarantees hold despite Byzantine behavior of a fraction
of the participants (nodes) maintaining the cryptocurrency service. 
To this end, cryptocurrencies rely on distributed ledger technology,
which serves as a transaction database, containing
the global history with all transactions. To build this global history, typically
consensus-based blockchain solutions are employed. 
A blockchain consists of a replicated linked list of immutable blocks, each block comprising
batches of transactions. This list is maintained by a
large number of nodes to tolerate malicious behavior of a
small group of nodes and still reach agreement on the blocks and their content
with a consensus protocol. An honest node will only propose and agree on blocks
that contain valid transactions: i.e., the transaction is properly signed by the
current owners of the funds, it has not been executed already, and 
and the senders' balance is high enough.\footnote{To facilitate validity checks, many 
cryptocurrencies require outgoing transactions to link to a previous incoming transaction. 
Thus, an attempt to double spend consists in getting the same transaction into multiple blocks that the receipients consider valid by mistake.}
Using such a blockchain, virtual currency can be transferred from
senders to receivers in a fully distributed manner, cutting out
any middle man or trusted third party. This feature has gained
enormous visibility and is envisioned to transform the financial
sector and potentially bring disruptive innovation to many
other sectors that traditionally rely on trusted third parties.

Bitcoin and other cryptocurrencies run on top of a \textbf{P2P network}.
Over this network, nodes send and receive blocks and transactions, which are the basic data structures of cryptocurrencies. 
In permissionless protocols, such as  Bitcoin and Ethereum, any machine can join the network and become a node of the P2P network. A node bootstraps its operation with a \emph{discovery protocol} to establish connections with other nodes in the system. 

In most cryptocurrencies there are two roles a node can assume: peer or miner.
Peers can create and send transactions. Peers verify the correctness of received transactions and blocks and relay them if valid. Miners do anything a peer does, but they also generate blocks.
Transactions and blocks are typically propagated in the network using a flooding or gossip protocol. E.g., when a node either creates or receives a  transaction or block in Bitcoin, it announce that item and may request it from peers if receiving an announcement of an item it does not have yet. 
A node does not forward invalid items. A node keeps valid items in memory (the mempool) and answer requests for them. This way, each node in the network will eventually learn about every new item.
The underlying wire protocol prescribes the data encoding and how to use which transport protocols. E.g., Bitcoin clients establish a TCP connection and perform a protocol-level three-way handshake informing each other of the height of the blockchain as they
know it and the software version they use~\cite{networkBitcoinit}. To support encryption and authentication, Ethereum defines the TCP-based DEVp2p protocol~\cite{wireprotocolehthereum}. 
After a handshake, all exchanged messages are encrypted and authenticated via key material
generated during the handshake. 

In a later of this paper, we provide a detailed discussion of block propagation and transaction propagation mechanisms and measurements. Furthermore, incentives to support reliable information forwarding and the different topologies of the virtual currency networks are presented. 

Blockchains are typically managed by a peer-to-peer network of nodes which collectively create and validate new blocks. 
\ifthenelse{\boolean{SHORT}}{}{In order to improve scalability,
additional payment channel networks may be implemented offchain,
offloading the blockchain. Such protocols rely on the ``parent blockchain" for security. }
Interestingly, however, cryptocurrency networks and their applications and usage, can differ significantly from traditional communication networks.
These differences influence the required performance, security,
and incentives, and touch all layers of the network stack.
Indeed, while early solutions relied on either a centralized issuer~\cite{rivest1997electronic,yang2003ppay}
or creating inter-user credit~\cite{fugger2004money}, 
which required users to trust the original issuer;
decentralized systems more critically depend on the \emph{network} 
to connect their constituent parts. One main challenge is the fact that cryptocurrencies are a relatively recent concept 
and many aspects are not well explored and documented, even compared to other components of blockchains and cryptocurrencies. 
With the exception of Bitcoin and Ethereum, many cryptocurrencies lack substantial documentation about their operational details, other than  
information scattered in the source code repositories.
At the same time, the dependability of cryptocurrencies is becoming increasingly important, and vulnerabilities and inefficiencies are a major concern given the corresponding direct financial implications.

\subsection{Characteristics}
Before diving into the details of the state-of-the-art and research 
challenges, we give an overview of some of the distinguished
characteristics of cryptocurrency networks. 

A first important aspect concerns \textbf{incentives}. Since there is no central party or 
consortium paying for the resources (bandwidth, CPU, storage, ..) necessary to maintain
the distributed ledger, the participating nodes must be remunerated through the 
protocol directly.  Similarly to traditional peer-to-peer networks,
nodes maintaining distributed ledgers require incentives to motivate nodes to propagate
information (transactions, blocks, control information) between them.
In addition, cryptocurrencies also need an incentive system to motivate nodes to verify blocks and the transactions included in them and discard invalid ones. 
In Bitcoin, nodes generating blocks are called miners. As a remuneration of their work the creator of block obtains a block reward and a fee for each transaction in the block.
Thus, miners have an incentive to keep the knowledge of any transaction that offers a high fee to themselves instead of forwarding them, as any
other node that becomes aware of the transaction will compete to include it in a block first and claim the associated fee. 
Additional incentive questions involve the use of lightweight nodes (know as SPV wallets) that rely on messages from a full-node for their operation. SPV wallets do not hold a complete copy of the blockchain and so must rely on other nodes to track payments sent to them. 

Another interesting characteristic is formed by the prevalent \textbf{communication pattern}.
Many cryptocurrency networks are characterized by frequent broadcast operations, e.g., related to the communication of transactions and 
states of the blockchain. Furthermore, systems such as Bitcoin do not follow a complex multihop routing scheme but employ a simple flooding-based strategy where all peers in the network replicate
the information that has been flowing through the system so far, i.e., keep a complete copy of the blockchain.
Hence, there is no need to forward queries to other peers, as all
information should to be available at a neighbor.

\ifthenelse{\boolean{SHORT}}{}{
There are also differences in \textbf{routing} itself.
For example, existing network routing algorithms for data transmission
experience unique challenges when applied, to payment channel networks.  
In payment channel networks, link
capacities represent payment balances, which can be highly dynamic:
messages are financial transactions, which may change liquidities and introduce
additional security requirements (e.g., related to privacy), 
and different routes may be used at different monetary costs (e.g., as intermediate
nodes charge fees for forwarding). If link capacities (representing funds) or ``liquidities''
should be kept private, it may become difficult to design an efficient route discover process \cite{saar-economy}.
}

The \textbf{network topology} in cryptocurrencies can be fairly different from traditional networks.
For example, while Bitcoin and Ethereum rely on flat random graph topologies, Cardano ~\cite{cardano-topology} uses different roles influencing how nodes connect to each other and to users (see Section~\ref{sec:topology}). 
\ifthenelse{\boolean{SHORT}}{}{
Also in this regard, offchain networks are particularly interesting. 
Since in these networks, capacities represent financial balances
which may need to be kept confidential, new threats may be introduced 
which are not encountered in classic P2P networks. 
}

In terms of \textbf{security} and dependability, cryptocurrencies critically depend on a correct functioning of the
consensus layer, and the knowledge of the set of information consensus is to be agreed on (e.g., blocks and
transactions). Flooding or gossip protocols are used for the
propagation of the required information to all peers of the
network. While unstructured P2P networks have been used
for decades (e.g., Gnutella) and were extensively analyzed, 
the considered adversarial models do not match well the threats to blockchain systems.
For example, anonymity providing networks (i.e., onion routing networks~\cite{goldschlag1999onion}) have different requirements regarding information propagation than blockchain based systems. Commonly considered requirements in anonymity providing networks are high performance, low bandwidth cost, resistance to traffic
analysis, and resistance to denial of service (DoS).
For example, distributed denial of service attacks can be used to gain advantages in mining, voting, and other business and
protocol-related activities~\cite{apostolaki2018sabre,johnson2014game,vasek2014empirical}.
 To prevent malicious nodes from flooding the network with invalid blocks, nodes use a store-and-forward propagation model, where each node downloads the full block and verifies it prior to propagating it to its peers. This model allows nodes to identify any node which propagates invalid blocks as malicious, and limits
the effect of such attacks to the nodes which are directly attacked.

There are also implications on \textbf{performance}.
While individual nodes may support high transaction rates, the distributed propagation can introduce novel kinds of bottlenecks. Indeed, one of the main issues in blockchain systems is their scalability. Increasing the number of transactions processed by the system naturally requires more resources such as bandwidth and storage, and especially the consensus protocols underlying cryptocurrencies can slow down execution, or even effect security. 
For example, Nakamoto's consensus protocol which relies on the longest chain for Bitcoin is known to be susceptible to attacks by weaker attackers as transaction throughputs increase. 
The main underlying cause of this decline in security is the fact that blocks containing more transactions propagate more slowly through the network, which causes forks to form in the blockchain~\cite{decker2013information}. As a result, a great deal of work has been devoted to improving block propagation times (see in Section~\ref{sec:blockpropagation}).
In order for Bitcoin to function as a decentralized system, it must allow nodes to receive
blocks at a higher rate than the rate at which blocks are being produced. Indeed, if blocks
are produced at a higher rate than a node is capable of receiving them, then said node
cannot keep track of balances stored in the blockchain, cannot determine whether or not
transactions and blocks are valid, and is in effect excluded from the Bitcoin network. The
block propagation time to reach the majority of the network does not depend solely on a receiving node's bandwidth, but also on the network topology, the bandwidth of all nodes, and the manner in which blocks propagate. 
\ifthenelse{\boolean{SHORT}}{}{
In offchain networks, as mentioned also earlier,
novel issues arise which lie at the intersection between performance and security. 
In offchain networks, to protect user privacy,
only the total capacity of a channel is disclosed, but not how the funds are distributed among the
the channel participants~\cite{silentwhispers,roos2017settling,sivaraman2018routing}. Channel transactions might therefore
fail and the routing algorithms attempt different execution
paths until one succeeds, which can introduce delays. Routing algorithms
in payment channels networks therefore,
have to account for the unique characteristic of channels to
provide satisfactory path recommendations. 
}


\section{Block Propagation}\label{sec:blockpropagation}

\subsection{What is it about?} \label{subsec:block_propagation_intro}
Blocks in cryptocurrency protocols are used to establish common state. They form the input the consensus protocol strives to reach agreement on. Blocks order transactions, thus the state of the network can be constructed by following the ordering of transactions included in blocks in the consensus chain. Transactions once included in a block deep in the consensus chain are considered confirmed. Therefore, block propagation is an issue of utmost importance to the consensus process. How fast miners learn about new blocks, and how quickly blocks can be created and validated are crucial for  the efficiency of a cryptocurrency.


A block consists of a header and a set of transactions. These transactions can be relayed by the sender together with the block, but this wastes bandwidth if they are 
already stored in the mempool of the receiver.

Blocks are typically re-propagated to all connected peers as soon as basic validity of the announcement has been established (e.g. after the proof-of-work check). In Bitcoin, propagation uses the NewBlock and NewBlockHashes messages. The NewBlock message includes the full block and is sent to a small fraction of connected peers (usually the square root of the total number of peers). All other peers are sent a NewBlockHashes message containing just the hash of the new block. Those peers may request the full block later if they fail to receive it from anyone within reasonable time.

Blocks can be relayed with a compressed encoding. Efficient propagation of blocks is critical to achieving consensus, reducing storage bloat, overcoming network firewall bottlenecks, and allowing scaling to a large number of transactions per second.
Delayed blocks can lead to forks~\cite{decker2013information}: 
based on measurements of the rate of information propagation in the network, 
the propagation delay in the network can be the primary cause for blockchain forks. blockchain forks should be avoided as they are symptomatic for inconsistencies among the replicas in the network. As a mitigation strategy, the authors propose pipelining the block's delivery, i.e., starting to transmit blocks before they have been fully validated.

One of the reasons for such delays is churn. Imtiaz et al.~\cite{imtiaz2019churn} report that almost all (97\%) Bitcoin nodes are connected intermittently only, which results in significant numbers of unsuccessful exchanges, roughly twice the figure for continuously connected nodes. In particular, they demonstrate experimentally that this churn leads to a 135\% average increase in block propagation time (i.e., 336.57 ms vs 142.62 ms), and can lead to as high as an 800-fold increase in the worst case measured.

Also permissioned blockchain networks based on Byzantine Fault Tolerance (BFT) consensus algorithms are highly affected by the propagation time. Nguyen et al.~\cite{nguyen2019impact} demonstrate how a network delay leads to a 30 times larger offset in the consensus layer in Hyperledger.

Often, miners collaborate in mining pools to share the risks and rewards of finding blocks. To this end, a dedicated server is connected to a node that acts as a gateway to a cryptocurrency network. This node gathers newly transmitted transactions
and newly built blocks to construct a new block template. The template header is then sent via a mining pool server to the miners which attempt to complete it to a valid block. In the simplest approach for Bitcoin, the miners try different values for the nonce field in the header. If the resulting hash has enough leading zeros for the current difficulty level, i.e., when the block is completed, it is sent back to the server, which then uses the gateway node to publish the newly formed block to the network and distributes the reward among the miners of the pool corresponding to their contributions.

In 2017, Bitcoin derived at least 95\% of its mining power from 10 mining pools; in the Ethereum network, 6 pools are responsible for 80\% of the mining power~\cite{luu2017smartpool}.

\subsection{State of the Art} \label{subsec:block_propagation_art}

\begin{figure}[t]
	\centering
	\includegraphics[width=0.8\columnwidth]{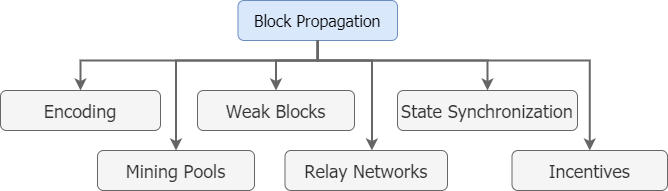}
	\caption{Block Propagation Topics}
	\vspace{-.4cm}
\end{figure}

A measurement campaign~\cite{neudeckercharacterization} showed that 2016-2018 it takes between 2 and 20 seconds until 90\% of the nodes announce the reception of a new block, with a tendency for shorter propagation latency in the more recent past. This is a vast improvement considering that Donet et al.~\cite{donet2014Bitcoin} reported that after 84 seconds new blocks have been reached 50\% of the nodes and less than 1\% of the blocks is known by 90\% of the nodes in the same time.

\noindent \textbf{Compressed block encoding.}
Many proposals to minimize the bandwidth consumption for block propagation exist. One such proposal for Bitcoin addresses the inefficiency of broadcasting blocks with all the transactions included. By the time a new block is created, it is very likely that most peers have these same transactions stored in their mempool. As such, relaying new blocks causes inbound bandwidth spikes for receiving nodes and potentially large outbound bandwidth spikes for nodes that receive blocks before their peers, since they will flood the network with the new, raw block data. 
	
Xtreme Thinblocks (XThin)~\cite{tschipper2016buip010}, 
deployed in Bitcoin Unlimited (BU) clients uses 
Bloom filter encoding the transaction IDs in nodes' mempool, thus only missing transactions must be exchanged. 
In an alternative, the Compact Blocks protocol~\cite{compactblocksrelay}, deployed in the Bitcoin
Core, Bitcoin ABC, and Bitcoin Unlimited clients, 
 the block’s transaction IDs are announced shortened to 6-bytes. If the receiver has missing transactions, it
requests them separately. 
\ifthenelse{\boolean{SHORT}}{}{
For $n$ denoting the number of transactions, the network cost is hence $6n$ bytes 
while, XThin’s cost is in the order of  $m \log f  + 6n$, with $m$ referring to the number of transactions at the receiver and a false positive rate of $f$.}
Thus, if the receiver is missing many transactions, Compact Blocks have an extra roundtrip time compared to Xthin, which may cost more if enough transactions are missing. 
Graphene~\cite{ozisik2019graphene} combines the use of a Bloom Filter with Invertible Bloom Lookup Tables (IBLTs) \cite{goodrich2011invertible}.
The main concept of Graphene's approach consists in shrinking the size of the sender’s Bloom filter by increasing its false positive rate, and correcting any false positives at the receiver with an IBLT. The summed size of the two structures is smaller than using either alone. In practice, this technique performs significantly better than Compact Blocks for all but the smallest number of transactions, and it performs better asymptotically than any approach relying on Bloom-filters only.
%
%
In comparison to XThin, Graphene uses significantly lower bandwidth both when the receiver is and is not missing transactions. However, Graphene may use an additional roundtrip time to repair missing transactions.
\ifthenelse{\boolean{SHORT}}{}{
The protocols described above rely on a single peer to send the complete data of a block, opposed to using multiple peers to transmit partial data. }
Velocity~\cite{chawla2019velocity} is an approach that exploits the fact that typically several peers already have (parts of the) data in a block. To this end, it applies Fountain codes, which provide a mechanism by which information can be encoded such that the resulting segments can be probabilistically re- assembled into the original data when the number of the received segments exceeds a threshold. 

\ifthenelse{\boolean{SHORT}}{}{
Upon receiving an inv message, a receiver requests any unknown blocks using a get\_sym request to all its neighboring peers. Peers which have information on the requested block(s) respond with repeated sym responses, each encoding one symbol. The receiver collects these symbols and reconstructs the corresponding blocks when it has received a sufficient number of symbols. If the reconstruction succeeds, it notifies its peers to stop symbol transmission. Note that this approach can be used for node bootstrapping in addition to speeding up synchronization.
}

\noindent 
\textbf{Stratum Mining Protocol.}
Stratum~{\cite{stratumBitcoinit, stratumslushpool} is the de-facto standard mining communication
protocol used by blockchain-based cryptocurrency systems. It enables miners to reliably and efficiently fetch jobs from mining pool servers. 
	
Stratum was initially a proposal for an open source client-server overlay protocol to support lightweight
clients. The Stratum mining protocol extended this proposal to a networking protocol for pooled mining services on the Bitcoin network and many other blockchain protocols. The protocol establishes client-server connections using plain TCP sockets between mining clients and a pool operator or server to distribute new work defined through a blockchain's proof of work protocol in human readable format.
	
	
Recabarren et al.~in~\cite{recabarren2017hardening}
exploit Stratum’s lack of encryption to develop passive and active attacks on Bitcoin’s mining protocol, with important implications on the privacy, security and even safety of mining equipment owners. 
Active attacks can hijack shares submitted by miners, and their associated payouts, by modifying TCP packet surreptitiously  without causing disconnections and session resets.
To mitigate such attacks, the authors proposes Bedrock, a Stratum extension that protects the privacy
and security of mining participants with mining cookies.  Each miner shares a secret with the pool and includes in its puzzle
computations, preventing attackers from hijacking the solutions.

\noindent \textbf{Weak blocks.}
In order to speed up block propagation even further, one approach is to let miners broadcast blocks they are working on before they have finished the corresponding proof of work. More precisely, so called \textit{weak} or \textit{near} blocks  whose proof of work is insufficient for the target difficulty, can be disseminated early. As a consequence, when the block is fully mined the corresponding payload has been received and validated by most nodes already and only the headers needs to be broadcast and processed~\cite{weakblockthoughts}.

Traditionally, the weak blocks are discarded in Bitcoin, wasting their proof of work entirely, contrary to the mission of securing the chain with any and all sufficient proofs of work. Weak blocks by definition have shorter interarrival times and can be used by miners to both receive strong confirmation signals for weak transactions (transactions in weak blocks) as well as anticipate forks sooner. Many updates to Nakamoto Consensus have been proposed that utilize similar ideas, yet no protocol change to utilize weak blocks has made its way into the Bitcoin Core source code. Some, such as BitcoinNG \cite{eyal2016Bitcoin}, exploitt weak blocks to store and propagate transactions. The \textit{key} blocks serve to elect a new leader, granting that miner the right to extend the chain with \textit{weak} blocks. The protocol splits rewards between miners of previous leader elections to incentivize against malicious behavior such as selfish mining or hidden block extension attacks. Another similar proposal termed \textit{Flux} \cite{zamyatin2018flux} augments the existing Bitcoin protocol with weak blocks such that chains of weak blocks or sub-chains contribute to a chain's proof of work. Using the heaviest chain rule as its consensus rule, it can provide faster transaction confirmation times by ensuring that \textit{key} blocks that link to sub-chains contain transactions included in the sub-chain's \textit{weak} or \textit{sub}-blocks. This can work in practice without the buy-in from all miners as well. One can imagine that if a certain number of miners opts in for broadcasting \textit{weak} blocks, \textit{key} blocks (which would also serve as legacy Bitcoin blocks) could ensure that the dominant chain contains some sub-chains of \textit{weak} blocks.

\noindent 
\textbf{Relay Networks.}
In parallel to the public P2P protocol, separate relay networks have been designed to increase network efficiency for miners. The first such system for Bitcoin, called Bitcoin relay network~\cite{Bitcoin-relay}, achieved this by disseminating blocks without full block verification and retransmitting known transactions. It consisted of a few nodes (supported by donations) scattered around
the globe, all of which peer with each other.  Another approach, Falcon \cite{falcon}, relies on cut-through routing for faster block propagation in addition to minimal validation and a hand-optimized topology. 
More recently, FIBRE~\cite{fibre} has been initiated to provide a similar service by combining cut-through routing with compact blocks and forward error correction over UDP (the normal Bitcoin protcol uses TCP) for registered users. 
Both Falcon and FIBRE can greatly reduce block propagation times and block orphan rates in the Bitcoin network, as shown in
~\cite{otsuki2019effects}.
However, it is important to note that neither was designed, nor is suitable, to scale Bitcoin. Bitcoin cannot rely on these relay networks to achieve higher throughput, since the use of a relay network to
scale, places the control over which transactions are included in the blockchain, and which miners may participate, in the hands of its operator. For example, the relay network operator may choose (or be coerced) to propagate blocks only from one group of miners, and reject all others, or to propagate blocks only to one group of miners, and not to others. Worse still, it might reject all blocks which contain transactions involving a specific address, effectively banning its owner from using it. 

\noindent 
\textbf{State Synchronization.} For new nodes to be able to contribute to the P2P network quickly, Ethereum provides a state synchronization protocol. 
The first message sent by two Ethereum peers after the handshake describes their status containing information on the node’s protocol version, network ID (multiple Ethereum networks
exist),  the hash of the genesis block, the best known block hash and the currently used difficulty. Only connections to nodes operating on the same network ID and genesis block are maintained. Based on their best block hashes the nodes will then synchronize their available information.

When a node joins the Ethereum
network, it obtains a local copy of the full blockchain
by first requesting
block headers, which include block meta information such as parent
block hash, miner address.
After it has compiled a list of missing block hashes, the node then
sends requests to retrieve full block contents
and verify the validity of the blockchain.

In Ethereum two validation mechanisms can be distinguished: 1) block
header validation and 2) blockchain state validation. Block header
validation, ensures that a block’s parent block hash, block number, timestamp, difficulty, gas limit, and valid proof-of-work hash are correct. In contrast, blockchain
state validation consists of sequentially executing all transactions and thus requires significantly more
computation and time.
In order to reduce the time for new nodes to synchronize and
validate the entire blockchain, the \emph{fast
sync mode} has been introduced. Instead of running blockchain state validation on all blocks since genesis (as necessary in Bitcoin), header validation is run until a pivot point block close to the most recent head of the blockchain is chosen (using GET\_RECEIPTS messages for meta information
including gas consumption, transaction logs, and status code). At
the pivot point, a fast sync node utilizes GET\_NODE\_DATA messages
to download a global state database at that block. From the pivot
point onward, the node performs full blockchain validation.

\noindent
\textbf{Incentives for Block Propagation.}
For cryptocurrencies to function properly, blocks need to be propagated promptly upon their creation to all other users in the network. This is crucial both to the liveness and to the security of these protocols. To this end, there is a need to examine that miners are incentivised to follow block propagation rules, and not to vary from them. Selfish mining attacks \cite{eyal2014majority, sapirshtein2016optimal} aim to increase the relative fraction of blocks mined by an attacker through timing the release of blocks created by an attacker. The strategies differ based on how long the attacker waits before publishing blocks from a secret chain. These attacks show that there are cases in which miners can profit by not propagating blocks as soon as they are created. 

Another case in which miners can profit by deviating from vanilla block propagation is \textbf{SPV mining.}
In a similar vein, the “SPV (simplified payment validation) mining” concept can decrease the block propagation latency, by avoiding the full verification of blocks and instead relaying them partly unchecked. Originally, the approach has been developed to expedite mining:
In order to build on top of the previous block and extend the chain, miners need the hash of the previous block. However, they do not need the full block with all the transaction data in order to start mining. It is in fact sufficient to only have the hash of the previous block header in order to mine a valid block. The incentive for SPV mining is a rush to mine blocks as fast as possible to increase profits. Waiting to download the full block and validate all of the transaction results in idle time which can accumulate to lost profits. Therefore miners may be tempted to find the next block before they have even had time to download and verify the previous block. Like this, miners avoid putting any transactions in the block (apart from the coinbase transaction that rewards the miner), since they cannot  know which transactions were in the previous block. Including transactions could result in double spending (which would deem the block invalid). SPV mining is one of the reasons that empty blocks appear on the blockchain~\cite{wang2019measurement}.
Moreover, SPV mining increases the probability of an invalid block being used to extend the chain and mine the next block linking to an invalid block (since the transactions are not validated by the following block, or even multiple blocks). This in turn results in the network being less reliable for payments as double spends are more likely.
In fact SPV mining has already caused a split in the network in the past: In 2015 there was a change implemented in the Bitcoin protocol (regarding enforcing BIP66 strict DER signatures) that was supposed to go into action after $95\%$ of the network updated their software. The way in which this was implemented is the following: Once 950 of the last 1,000 blocks were version 3 (v3) blocks, all upgraded miners would reject version 2 (v2) blocks.
On 4 July 2015, shortly after the threshold was reached, a small miner (part of the non-upgraded $5\%$) mined an invalid block. Unfortunately, it turned out that roughly half the network hash rate was mining without fully validating blocks, and built new blocks on top of that invalid block, causing a split.

SPV mining is also an issue in the Ethereum, and in 2018 it was discovered that F2Pool (one of the largest mining pools at the time) was engaged in SPV mining (creating a dispropotionate amount of empty blocks. \footnote{\url{https://medium.com/@ASvanevik/why-all-these-empty-Ethereum-blocks-666acbbf002}}

Another similar strand of attack is Spy mining among competing mining pools. Spy mining occurs when attackers join the pools of others to obtain hints about new blocks appearing on the network. When a spy detects such information via the changed headers sent to it in the Stratum protocol, it can notify its other pool to avoid wasted work. Thanks to SPV, the miners can start mining a new block without seeing the old blocks content.

\subsection{Open Questions} \label{subsec:block_propagation_questions}

Latency and throughput of crypto currencies depend on the efficiency of block propagation. Therefore, mechanisms and incentives to spread newly minted blocks as quickly as possible while minimizing bandwidth waste are crucial.  At the same time, the system design must not forego safety and the protocols in place must ensure that blocks contain valid transactions despite the hunt for speed.

\noindent\rule{1\linewidth}{1pt}
\noindent \begin{openq}
\noindent\textbf{How to accelerate block propagation?}
	Several approaches to optimize the exchange of information in blocks between two nodes as well the dissemination in the network have been presented. Yet, both dimensions provide opportunities for empirical studies  of the status quo and subsequently to devise new approaches to overcome this barrier to faster transaction rates.
\end{openq}

\begin{openq}
\textbf{How to incentivise mining blocks with many valid transactions?}
	SPV mining still poses risks, thus designing a system preventing this behavior, e.g., with economic incentives is still open. 
\end{openq}

\begin{openq}
\textbf{How to broadcast blocks within a pool?}
	Pools which are very well connected internally and therefore can disseminate newly minted blocks fast, have an advantage in mining over their competitors. Within a pool one could consider a more permissioned model with a more structured overlay topology for speed, balancing the possible velocity gains with the risk of new attacks.
\end{openq}

\begin{openq}\textbf{Efficient network design for mining pools:}
How to get headers to miners quickly, how to disseminate the
accomplished work in whole network?
\end{openq}

\begin{openq}\textbf{Should pools be “allowed”?} 
If not can they be prevented? What tools can be used? Can pools be monitored? 
There are also questions related to incentives 
(e.g., to force miners to have copies of the blockchain to make pools less attractive).
\end{openq}
\vspace{-.2cm}
\noindent\rule{1\linewidth}{1pt}

\section{Transaction Propagation}\label{subsec:transaction_propagation}

\subsection{What it is about?}\label{subsec:transaction_propagation_intro}
One of the main services provided to users in cryptocurrencies is the propagation of their transactions. Users' transactions must reach miners in order to be included in blocks, and similarly miners are interested in obtaining transactions of users in order to be able to collect their associated fees. Hence, delays in transaction propagation result in possible delays for transaction confirmation for users, and may cause losses of funds for miners. 

Bitcoin’s method for propagating transactions is flooding-based, using the same underlying P2P network used to relay blocks. Thus, if a user sends a transaction to a node, it is potentially sent to all other peers of that node, and then propagated onwards. 
Transaction relaying has been measured to be slower than block propagation~\cite{donet2014Bitcoin}. While 50\% of blocks were broadcast to 25\% of the nodes in less than 22 seconds, 17 minutes are needed to relay 50\% of the transactions to the 25\% of the nodes in the sample.
Similarly, Neudecker~\cite{neudeckercharacterization} reports that since 2017 transactions propagate to 50\% of all nodes within around 5 seconds and to 90\% of all nodes within around 15 seconds, both with a trend to increase over the horizon of the measurement campaign.

Kim et al.\cite{kim2018measuring} observed that transactions dominate the network traffic, with clients using different strategies for transaction dissemination. Different  protocols have been suggested, such as Geth which relays transactions to all peers, and Parity which forwards them only to a subset of size square root of number of peers in the network.

\subsection{State of the Art}\label{subsec:transaction_propagation_art}

\noindent \textbf{Amplified DDoS.}
One main concern with respect to transaction propagation is that attackers will try to send many transactions to nodes. If each such transaction is later propagated to the entire network unconditionally, then attackers would be able to amplify any large scale DoS attack---sending a single message causes many more messages to be generated throughout the network. Thus, an attempt is made to charge the attacker for sending messages. Unlike with blocks, that are rate-limitted by the very fact that they require a proof-of-work in order to be valid, transactions bear no such limits. Their main cost for senders are reflected in the fees paid by the sender. These fees are not guaranteed. In fact, if a transaction is propagated to the entire network and later is not entered to the blockchain, its fees are not collected. Thus, in Bitcoin, miners only convey transactions that enter their mempool -- these in turn are transactions that pay a sufficient fee that is likely to lead to their inclusion in a block.

Similar concerns apply when users wish to increase the fees of their transactions (possibly after finding out the the fee is insufficient to enter a block quickly enough), or to users who wish to double-spend previous un-confirmed transactions. If users are allowed to replace transactions with very modest fee increases, and the new transactions are propagated everywhere, again an attack is possible: attackers will simply send one transaction and then replace it with newer transactions with insignificant fee increases, thereby flooding the network once again. Thus, miners typically set a minimal fee increase to replace a transaction in the mempool and have it re-sent.

Similarly, if a user double spends a transaction and redirects its funds elsewhere, miners will not typically relay the double spend. This serves both as a countermeasure for transaction flooding and as a tool to support 0-confirmation recipient policies (which we discuss below).

\noindent 
\textbf{Information Eclipsing and 0-Confirmation policies.}
Some users may not wish to wait until transactions are included in blocks. They then can adopt a somewhat risky policy of accepting 0-confirmation transactions, i.e., they can consider funds received if a transaction is propagated through the network (with fees that they estimate are sufficient to enter a block).  
Since these transactions are not yet included in a block, they are vulnerable to double spends. Most notably, 
double spends by miners that need only a single block that contains a conflicting transaction before the double spent transaction is included in a block itself (known as Finney attacks).

\begin{figure}[t]
	\centering
	\includegraphics[width=0.9\columnwidth]{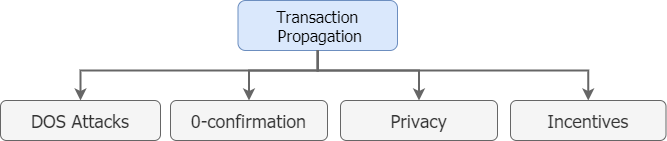}
	\caption{Transaction Propagation Topics}
	\label{figure::sec_transaction_propagation}
\end{figure}

Due to the policy of miners mentioned above, double-spend transactions are not entered into the mempool of nodes and are not further propagated to others. This implies that some nodes are essentially ``eclipsed'' by their neighbors and may never find out about the existence of a double spend, simply because these neighbors first received one transaction (and forwarded that one to the node), but did not relay a conflicting transaction that was later received, leaving the node unaware of its existence. Information eclipsing hinders nodes that would like to accept 0-confirmation policies.

Still, such attacks require mining power which can present some barrier to the attack. The question of how easy it is to attack 0-confirmation transactions without holding mining power naturally arises. 
Karame et al. \cite{Karame:2012:DFP:2382196.2382292} analyzed such attacks. They discuss countermeasures such as waiting a while before accepting the payment (and checking to see if conflicting transactions are propagated) and judge that these are not always sufficient to prevent attacks. They propose modifying protocol rules so that nodes forward double spending transactions instead of dropping them to avoid information eclipsing. They however do not analyze the effect of denial of service attacks that may be aggravated as a result. 

\noindent 
\textbf{Privacy.}
An active attacker may wish to identify the node from which transactions originated. By actively connecting to several nodes, it is possible that a curious observer will observe the transaction origin, or will be able to deduce it.

Biryukov et al. \cite{biryukov2014deanonymisation} present a deanonymization method for a significant fraction of Bitcoin users that correlates their pseudonyms with public IP addresses. The method explicitly targets peers behind NAT or firewalls, and can differentiate between nodes with the same public IP. They show ways to counteract the fact that nodes may use TOR to hide their IP and essentially utilize an anti DoS countermeasure in Bitcoin to cut off access to TOR. The main deanonimization technique is to identify nodes via the set of nodes it connects to (its entry nodes), transactions are mapped to a set of entry nodes, which uses a fingerprint to associate together transactions with similar sets, which suggests they belong to the same clients. The authors primarily propose to change the set of entry nodes often to avoid such correlations. 

Neudecker and Hartenstein evaluate a form of deanonymization in \cite{neudecker2017could}. They compare clustering approaches using transaction data (like signatures for different public keys that appear together in transactions) with clustering based on network data. Finding a correlation between the two likely means that both approaches in fact approximate the desired outcome. They show that for the majority of users no correlation between network information and the clustering performed on blockchain data could be found. A small number of participants do exhibit correlations that might make them susceptible to network based deanonymization attacks.

The Dandelion protocol \cite{bojja2017dandelion} is a suggestion to provide better network-layer anonymity for transacting users. 
The main construction in Dandelion is based on having a forwarding phase for transactions before they are widely distributed, trying to mask the origin of transactions. Nodes agree on some full ordering of the network (i.e. some Hamiltonian cycle) which is changed every few minutes in order to avoid the adversary learning it. Transactions are first broadcast along this path for a random (small) number of steps. After this phase, broadcast is done through diffusion as it is done today in Bitcoin. 
Dandelion++ \cite{fanti2018dandelion++} extends Dandelion to defend against adversaries that are allowed to disobey the protocol. The core improvement is moving from forwarding through a long line graph to a referral $4$-regular graph in the initial phase (before diffusion).


\noindent 
\textbf{Incentives.} 
The incentive of a node to participate in the dissemination of transactions is unclear.
As an example, consider the case of a large transaction with high fee. In Bitcoin, the miner's incentive is to not propagate the transaction to other miners, in order to reduce competition (so it can include the transaction in one of its own blocks and claim the fee).

Although this field was studied thoroughly in the context of P2P networks (examples in \cite{rapid-information-propagation}, \cite{sybil-proof-marketing}), one of the first paper examining this for cryptocurrencies was \cite{babaioff2012Bitcoin}. In this paper, the authors offer to augment the protocol with a scheme that rewards information propagation, while balancing it with the incentive to decrease competition. They show that their scheme is sybil-proof (robust against creating clones) and has low overhead (a total reward that is not too high).

An improvements is proposed in \cite{ersoy2018transaction}, where additionally to awarding the propagating nodes, the authors propose ``smart routing". In this mechanism, nodes directly route the transactions to a round leader, which is known in advance (first-leader-then-block (FLTB) consensus protocols). This mechanism increases the bandwidth efficiency by reducing the propagation of redundant transactions.

\subsection{Open Questions}\label{subsec:transaction_propagation_questions}

If the miners do not have access to a large number of transactions to put into blocks, the throughput and latency of the cryptocurrency will be suboptimal. In addition to the technical limits of spreading transactions widely, incentives play a very important role here and must be designed carefully.

\noindent\rule{1\linewidth}{1pt}
\begin{openq}
		\textbf{How to avoid transaction floods?}
	An attacker may try to flood the network with meaningless transactions and thus cause the nodes to waste resources on them. How can one quickly identify bad transactions and discard them? How to trade off the verification cost, punishment mechanisms and velocity?
\end{openq}

\begin{openq}
		\textbf{How to balance DOS prevention and 0-confirmation requirement?}
	On the one hand, nodes strive to avoid DoS attacks by not allowing the propagation of double-spend transactions. On the other hand, users want low-latency cryptocurrency systems and thus favor 0-confirmation policies. Mechanisms to meet these two conflicting goals would allow for a better user experience.
\end{openq}

\begin{openq}
		\textbf{How to model and analyse cryptocurrencies?}
	The costs and benefits of propagating the different protocol messages deserves a more thorough analysis under realistic utility assumptions. This will uncover further weaknesses of current mechanisms and inform the development of superior approaches.
\end{openq}

\begin{openq}
		\textbf{How to implement a suitable reward system?}
	Currently prevalent mechanisms incentivise nodes to keep transactions with high fees to themselves, instead of propagating them widely. To alleviate this shortcoming, one must find better methods to align the incentives of nodes and cryptocurrency users. In addition, strategies that allow a superior method to be adopted quickly
	must be developed.
	\end{openq}
\vspace{-.2cm}
\noindent\rule{1\linewidth}{1pt}

\section{Topology of the P2P network}\label{sec:topology}

\subsection{What is it about?}\label{subsec:P2P_topology_intro}
The Bitcoin P2P network topology is formed by each peer connecting to 8 nodes (outbound connections) and accepting up to 125 in-coming connections. Outbound destinations are randomly selected among known identities. 
In other cryptocurrencies, nodes are assigned roles that influence the topology. For example, Cardano distinguishes between mutually exclusive core, relay, and edge node roles~\cite{cardano-topology}. The core nodes create blocks and  run the consensus protocol, in other words, the core nodes maintain the blockchain. Relay nodes protect the core nodes, serving as intermediaries between the public internet where the edge nodes reside and the core node. If relay nodes are attacked, this may lead to a service interruption, but the integrity of the core nodes (and thus the Cardano blockchain) is not compromised. Relay nodes are fully under the control of the federated committee of initial Cardano stakeholders. Edge nodes create the payload of the blocks. They can be run by anyone on their computer to create currency transactions.  They cannot directly communicate with core nodes, only with relay nodes and with other edge nodes.
Also Ripple distinguishes between different roles for nodes, partitioning them in to superpeers and leafs\cite{ripple-overlay}. A nodes in the leaf role does not route messages and only connects to super peers. In the superpeer role, a peer accepts incoming connections from other leaves and superpeers up to the configured slot limit. It also routes messages. 

Before being able to send and receive protocol messages a node has to find other nodes to connect to join the network. This \emph{discovery} process typically relies on static information sources and/or distributed hash table (DHT) approaches, discussed in the first part of this section.
Knowledge of the network topology can give parties an advantage in the dispersal of information (blocks, transactions) which can lead to security risks. Because of this, there has been extensive research and development of tools and techniques aimed at exploring and mapping the Bitcoin P2P topology. The same holds true for many other crypto-currencies. We present here some of these works and their main contributions.

\subsection{State of the Art}\label{subsec:P2P_topology_art}

\noindent\textbf{Discovery.}
To join a cryptocurrency P2P network, a new node must find other nodes to connect with. In Bitcoin, a node first tries to connect to nodes it knows from participating previously. If no connections can be established this way, or if the node connects for the very first time, it queries a list of well known DNS seeds. As a last resource, it will try to connect to hardcoded seed nodes.
The DNS seeds are maintained by Bitcoin community members: some of them provide dynamic DNS seed servers which automatically get IP addresses of active nodes by scanning the network; others provide static DNS seeds that are updated manually and are more likely to provide IP addresses for inactive nodes~\cite{Bitcoin-p2p-guide}.
With the first messages exchanged between new peers, 
they inform each other of a random subset of locally known addresses with a timestamp of at most 3 hours ago. With this mechanism a list of addresses is maintained at each node. Each node will also accept incoming connections (up to 125 by default). The address of a new node is propagated through the network, 
so all peers can learn about it eventually. 

Ethereum and Cardano use mechanisms inspired by Kademlia~\cite{maymounkov2002kademlia}, a Distributed Hashtable (DHT) approach that has already been widely used for file sharing.
Kademlia assigns key-value pairs to sets of peers based on the distance between the key ID and the node IDs and a routing table structure is maintained that allows to find responsible nodes recursively in a logarithmic number of steps.

\begin{figure}[t]
	\centering
	\vspace{-.2cm}
	\includegraphics[width=1.02\columnwidth]{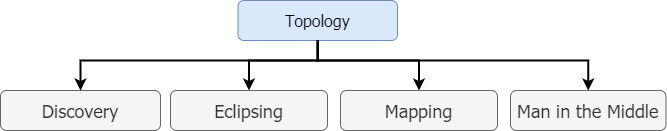}
	\vspace{-.3cm}
	\caption{Topology Topics}
	\label{figure::sec_topology}
	\vspace{-.5cm}
\end{figure}

While Kademlia and its derivatives have been used for many years, 
there is no formal proof for its performance and robustness.  
More theory-oriented approaches~\cite{bortnikov2009brahms,guerraoui2013highly,young2013towards} on the other hand, have not established themselves as alternatives.
\cite{guerraoui2013highly} shows how to maintain clusters of size O($\log n$), each containing more than two thirds of honest nodes with high probability. Even when the system size can vary polynomially with respect to its initial size, the
communication cost induced by each node arrival or departure is O(polylog $n$). The approach guarantees robustness to a Byzantine adversary controlling a fraction 1/3 of the nodes (could be 1/2 with the application of cryptography). 
The proofs guarantee polylogarithmic maintenance and sampling overhead and rely on assumptions  of a synchronous network. 
Alternatives, e.g., \cite{young2013towards}, based on  the BLS threshold signature scheme, demonstrate how a DHT with quorums of logarithmic size, the time and message complexity is polylogarithmic if up to a third of nodes per quorum is malicious. To this end,  any request needs quorum approval before getting answered or continued to avoid SPAM and DOS attacks and prevent wrong responses.
To remain robust despite churn, routing tables are maintained according to the Cuckoo Rule. Compared to other approaches, this method requires the creation and verification of many signatures, which is demanding. 


\ifthenelse{\boolean{SHORT}}{}{
If one can accept probabilistic failures, even more efficient options are possible.  Jaiyeola et al. describe in \cite{jaiyeola2018tiny} how to use
quorums sizes of O($\log \log n$), despite an adversary that controls a constant fraction of the computational resources in the network. Using their DHT approach,
all but an o(1)-fraction of the machines can communicate with all but an o(1)-fraction of the machines in the network in O($\log n$ poly($\log \log n$)) steps.
}
Instead of a DHT-based approach, Brahms \cite{bortnikov2009brahms} proposes churn and Byzantine resistant sampling. The paper presents an attack-resilient gossip-based membership protocol and shows how to extract independent uniformly random node samples from the stream of node ids gossiped by the first. It draws its power from an assumption of limited bandwidth avaialability: Byzantine nodes cannot send messages unlimitedly, if one node sends more often than expected, it is ignored. In Brahms unsynchronized gossip rounds, nodes send addresses they know to some other nodes (to reinforce the knowledge for underrepresented nodes), and request known addresses from other nodes (to spread existing knowledge).  If more than the expected  number of address are received in a round, Brahms does not update its view in that round to prevent malicious influence. Furthermore, locally known history also influences next views to avoid poisoning. Together with a sampling algorithm this ensures that nodes have an approximately uniform sample of the nodes in the system, even as long as every joining correct node knows some correct other node. Brahms offers a tradeoff between communication and storage and can thus be adapted to different needs. Such a sampling approach can also be run on top of other solutions.

\noindent\textbf{Mapping the network.}
In	\cite{ben2018vivisecting}, Ben Mariem et al. characterize and analyze Bitcoins' P2P network topology and main properties with a  purely network measurements-based approach. They present a BTC crawler in order to discover and track all active nodes of the BTC P2P network. They also discuss a passive approach to reconstruct the topology of a blockchain P2P network, which can unveil miners. \ifthenelse{\boolean{SHORT}}{}{The paper further describes the structure of addresses (IPv6, IPv4, Tor hidden addresses), and presents empirical results on the geo-location of active nodes around the world.}
Another approach is presented in~\cite{delgado2018txprobe}, uncovering the topology of the Bitcion network by processing orphaned and conflicting transactions 
More precisely, by sending different double-spending transactions one can determine if these nodes are connected. This is possible because of how a node's mempool maintains orphaned transactions and processes conflicting transactions. Their techniques allow mapping the topology of the Bitcoin network in roughly 8 hours.
The technique enables accurate topology reconstruction because of how nodes deal with both new transactions and orphaned transactions. \ifthenelse{\boolean{SHORT}}{}{Nodes that receive an inv-request and do not possess such a transaction will request it from the same peer with a get$\_$data message. However, if the node already possesses the transaction it will not send a get$\_$data message. By sending different nodes orphaned and conflicting, double-spend transactions and timing these actions correctly, an agent can identify if two of its peers are connected themselves. }
%
In \cite{miller2015discovering}, Miller et al.  determine the 
Bitcoin topology by utilizing the update method of the timestamp field in the addrMan of nodes in the network to learn about the topology. With this approach, they show that the deployed Bitcoin topology does not resemble a random graph. 

It is also possible to exploit transaction accumulation to map the Bitcoin network: In the two mechanisms presented in \cite{grundmann2018exploiting}, (i)  a node accumulates input transactions before propagating them, and then propagates them in the original order; (ii) nodes do not propagate contradicting transactions, so by propagating a contradiction to two nodes, they can tell which node is closer to the target (the one which got it first). \ifthenelse{\boolean{SHORT}}{}{By observing the INV messages from the nodes (the messages that propagate the transactions), the authors argue that it is possible to map the neighbors of a node (by sending different messages to different clients). The paper also presents
effectiveness and precision plots based on simulations.}

Similarly, other networks have been analyzed, including Ethereum and Monero.
\ifthenelse{\boolean{SHORT}}{\cite{kim2018measuring} unveils that in Ethereum}{
For a measurement study of the Ethereum network peers \cite{kim2018measuring}, 
researchers developed a special node that listens to the propagation mechanism, connecting it to the main network. The study finds that $48\%$ of all (3 million) discovered nodes do not contribute to the Ethereum Mainnet (do not run the Ethereum sub-protocol or do not operate on the main blockchain). The authors also show that $76\%$ use the golang implementation, $17\%$ rust, $5\%$ javascript (which might be cryptojacking web clients). The study also reports on the versions of the nodes (examining stable versions and up-to-date versions), the network size (activate peers), geographic distribution and node age. 
The authors observe that} more than 40\% (12\%) of the nodes use US (Chinese) IP addresses. 
The most widely used ASs of the nodes can be assigned to cloud hosting providers, 
dominating residential or commercial operators.
In  \cite{cao2019exploring}, a tools is presented to collect Monero’s P2P network information, including its network size, (geo-)distribution, and connectivity. The researchers set up $4$ Monero nodes across the world and found that $87\%$ of the nodes have a degree smaller than 8 (approx. $17\%$ of the overall number of edges), being able to map all outgoing connections of $99\%$ nodes in the network. The authors also succeeded to connect to $85\%$ of the nodes that they discovered, which may enable network level attacks (eclipse, BGP hijacking, DOS).

There has also been some work that aims to fix some of the
mapping issues raised. One approach is to secure blockchain 
network communication using SCION \cite{vorkapic2018secure}:
the design and use of a cryptocurrency network such as Bitcoin and Ethereum 
 The goal of using SCION is to facilitate client node communication with a more direct, point-to-point infrastructure to prevent attacks and numerous cases of lost cryptocurrency from network hijacks and BGP route poisoning/spoofing.

\noindent \textbf{Eclipsing / Splitting the network.}
Splitting the Bitcoin network can have severe consequences, as the shorter chain produced will not survive and as a result, many transactions are rolled back (and potentially double-spent), the revenue of miners from these blocks is lost. Partitions affect the ability of participants to operate on transactions. 
This may cause exchanges to stop receiving and sending the cryptocurrency, and merchants to be unable to get paid. There are several ways to isolate nodes in the network: for example, by disrupting the routing of traffic between them, or by causing nodes to connect only to attackers (Eclipsing).
We list here a few well-known eclipse attacks, and some techniques suggested in the literature to detect and avoid them.

In Bitcoin, nodes choose their peers from a list of stored IP addresses. This list is limited in size and IPs must be evicted if fresh ones are placed inside. IPs are placed in the list pseudo-randomly in a way that is based on the IP itself and the IP of the advertising node.
In \cite{heilman2015eclipse}, the authors explore ways of isolating nodes in Bitcoin by affecting the way that nodes choose their peers. The main idea of the attack is to announce many IPs to the node that are either controlled by the attacker or that have no node behind them. The node eventually evicts all IPs of honest nodes from the list, and will only connect to the attacker. 
The idea is to cause collisions in the placement of similar IPs and of IPs advertised by the same node, thus the attack above needs some minimal number of IP addresses controlled by the attacker to succeed. This number is not high in practice.  Also Ethereum is the target of eclipse attack constructions. In \cite{wust2016Ethereum}, Ethereum's peer-to-peer network  is partitioned without monopolizing the connections of the victim, which is possible due to the block propagation design of Ethereum. In this attack, the attacker can potentially keep the victim from receiving a block almost indefinitely. This attack could be used as an infrastructure for a double spend attack.
The authors also present an exploit that can force a node to accept a longer chain with lower total difficulty than the main chain (also using the block sync mechanism). In this attack a node that newly connects to the network and receives a chain that is longer than the valid chain but has a lower total difficulty because the adversary advertised a higher total difficulty than honest nodes. The attacker is therefore disconnected from the network.
The authors highlight a bug in Ethereum’s difficulty calculation as well. This can be used in an attack that prevents the victim from synchronizing with the valid chain. The paper also outlines countermeasures.

Two attacks on Bitcoin exploiting the networking stack are presented in~\cite{apostolaki2017hijacking}. First, BGP hijacking that is used to disconnect parts of the network. The network is shown to be poorly distributed, so that relatively few prefixes need to be hijacked in order to partition miners from each other. Once a partition is fixed, natural churn allows nodes to connect across the partition and blocks are once again propagated. 
A second attack proposed in the paper utilizes the fact that Bitcoin traffic at the time was un-encrypted. Intervention in the content of announcements of new blocks and transactions as well as requests for the data of recently announced blocks was shown to severely delay block propagation. As a result nodes are left uninformed of the latest blocks in the chain for longer periods, which causes miners to waste time mining blocks that will be discarded and will yield no reward, and users to be unaware of funds they may already have received. 

A game theoretic approach is used in \cite{tochner2018pick} to manage the list of known peers. Consequently, attackers need to corrupt a large number of nodes to eclipse a node successfully. Similarly to the Bitcoin protocol, the paper assumes that acquiring IPs from the same prefix is cheaper than acquiring the same number of IPs from multiple prefixes, and utilizes this fact in the peer selection mechanism to increase the attack cost.

In \cite{nayak2016stubborn} the authors investigate new ``Stubborn Mining'' attacks which combine eclipse attack with selfish mining \cite{eyal2018majority} attacks. In this work, the authors consider the same model against users who are also eclipsed in the network and show the effect to which eclipsed users help a stubborn mining attacker. 
Overall, eclipse attacks empower adversarial agents with a larger strategy space to continue running attacks, and when paired with stubborn mining strategies, enable an attacker to improve their relation fraction of block rewards beyond traditional selfish mining strategies. 

In \cite{tran2020stealthier}, Tran et al. present the EREBUS attack, that partitions the Bitcoin network without any routing manipulations, which makes the attack undetectable (even against bug fixes specifically adressing partitioning attacks). Adversaries who may control large transit ISPs, are able to mount the attack. The adversary utilizes a large number of network addresses reliably over an extended period of time. 
A fix attempt to this attack is suggested \href{https://github.com/Bitcoin/Bitcoin/pull/16702}{here}. This work enables Bitcoin core to prefer to connect to peers which are on different source ASNs to try to reduce the probability of any single host/path/hijack is relied on by a peer. The authors focus on the process of building the AS map, including simple filtering suggestions such as treating prefixes only reachable via a common upstream as if they were hosted directly on that upstream (by pulling routing information don diverse sources).

SABRE \cite{apostolaki2018sabre} presents a secure and scalable Bitcoin relay network resilient to routing attacks, designed to run alongside the existing peer-to-peer network and is easily deployable. \ifthenelse{\boolean{SHORT}}{}{The network is designed to efficiently handle high bandwidth loads, including Denial of Service attacks.
The relay network provides security to Bitcoin clients by enabling them to learn the latest mined blocks and to propagate them network-wide.}
The authors use properties of BGP to predict where would be a good place to host relay nodes --  locations that are inherently protected against routing attacks and on paths that are economically-preferred by the majority of Bitcoin clients. In addition, they provide resiliency through soft/hardware co-design through the use of caching, and offloading most operations to hardware (programmable network devices). This enables SABRE relay nodes to sustain load even when originating by DDoS attackers.

The effectiveness of mining pools can also be hampered by Distributed Denial of Service Attacks (DDoS) in order to disrupt their operations. As a consequence a competing mining pool is slowed down giving an advantage to other pools. This in turn may encourage individual miners to leave unreliable mining pools and join the attacker's pool as a result.
After currency exchanges, Bitcoin mining pools are the most frequent victim of DDOS attacks~\cite{vasek2014empirical}. Of 49 mining pools, 12 experienced (repeated) DDoS attacks. Based on a game theoretic model where pools can select between investing funds into additional mining equipment or DDOS attacks~\cite{johnson2014game},  larger mining pools have a slightly greater incentive to attack than smaller mining pools.

\noindent\textbf{Man in the middle attacks.}
In \cite{ekparinya2018impact}, the authors study the impact of Man-In-The-Middle Attacks on Ethereum. The paper looks closely at the feasibility of MITM and double spending attacks on simulated network corresponding to the real Ethereum topology with real network components. They show the impact of such attacks,
also gathering public information about the network, and mimicing the structure of its biggest 10 mining pools connected through 5 BGP routers, and performing BGP hijacking and ARP spoofing.
The authors find that the attack is almost infeasible in the public context (because of its structure), but in the case of route hijacking (e.g.  if Ethereum is deployed over a WAN in a consortium environment, and an adversary that has control on the border gate) could double-spend through either BGP hijacking or ARP spoofing with a success rate up-to $80\%$.

\ifthenelse{\boolean{SHORT}}{}{
\noindent\textbf{Empirical results.}
The open nature of the Bitcoin and Ethereum network attracted a number of studies of their characteristics. Donet et al.~\cite{donet2014Bitcoin} report  that in 2014 nodes placed in Unites States and China summed up to 37\% of the discovered nodes. Germany, United Kingdom, and Russia also had a large share of nodes of the network, with 9\%, 4\%, and 7\%, respectively. Japan, Brazil, Mexico, and China had low adoption rates, with the number of Bitcoin nodes being less than 3 per every 100k Internet Users. In contrast, the Netherlands, Norway, Finland, and the Czech Republic have the highest adoption rates, more than 10 times higher than those showed by Brazil. Most of the detected nodes, remain connected for a short time only,  merely 5,769 nodes remain (which represents only a 0.66\% of the discovered ones) after 37 days.

Neudecker~\cite{neudeckercharacterization} describes the methodology and results of his ongoing measurement campaign (2015-2018), including a comparison with measurements from other sources (all reproduce the same general trends, and show the same short-term effects).
The total number of connections studied varied between less than 6,000 connections in late 2016 and around 14,000 connections in 2018. The number of Sybil peers detected is generally very low (less than 50 prior to July 2017, less than 200 after August 2017), with the exception of short events in June 2017 and August 2017. The average share of peers connected for at least one day varies between 55\% and 75\%, the share of peers connected for at least one week varies between 20 \% and 50 \%.

In 2017, \cite{gencer2018decentralization} investigated bandwidth, latency, and decentralization metrics in the Bitcoin and Ethereum network. The observed bandwidth varies between different protocols, and the paper reports averages between 70 and 90 Mbps for Bitcoin and 55.0 Mbps for Ethereum nodes. The latencies they measure are 135ms and 171ms on average for Bitcoin and Ethereum nodes respectively. With respect to mining power, the authors show that in Bitcoin, the weekly mining power of a single entity has never exceeded 21\% of the overall power.
 In contrast, the top Ethereum miner
has never had less than 17\% of the mining power. Moreover, the top four Bitcoin
miners have more than 53\% of the average mining power. On average, 61\% of
the weekly power was shared by only three Ethereum miners. These observations
suggest a slightly more centralized mining process in Ethereum.

Feld et al.~\cite{feld2014analyzing} and Apostolokai et al. \cite{apostolaki2017hijacking} pointed out a strong AS-level centralization that may impact Bitcoin
network’s connectivity – i.e. 10 ASes contain over 30\% of peers,  with 39 IP prefixes hosting half of the overall mining power. 
}

\subsection{Open Questions}\label{subsec:P2P_topology_questions}

\noindent\rule{1\linewidth}{1pt}
\vspace{-.5cm}

\begin{openq}
		\textbf{Topology information hiding:}
How can the overlay topology be efficient yet make it hard for attackers to learn it and mount eclipse and hijacking attacks?
Answers to this question provide discovery mechanisms that strike a balance between containing truthful information and DOS resistance, e.g., using overlay rotation and sharding mechanisms 
that minimize the information necessary to participate in a crypto currency network.
\end{openq}

\begin{openq}
		\textbf{Giant honest component:}
Most crypto-currencies employ a flooding-based strategy to broadcast on a topology constructed with a (pseudo) random process. Non-honest nodes may choose to drop messages or forward outdated and wrong information. For a broadcast to succeed, thus enough honest nodes must be connected to each other via at least one path, not containing bad nodes. To this end, graphs must provide large connected components that consist of only honest nodes.
\end{openq}

\begin{openq}
		\textbf{Link failure models:}
Traditional failure models consider the number of faulty nodes as the main parameter when analyzing P2P networks. For a more nuanced analysis, link failures must be taken into account as well. Link failures can be modelled as random processes or in a worst-case fashion that is bounded in some way (e.g., a strongly connected union of available links when considering a time interval~\cite{haeupler2016analyzing}). More granular models and analyses tailored to the cryptocurrency conditions and constraints are necessary to better understand current cryptocurrency networks and to build the basis for future designs.
\end{openq}
\vspace{-.2cm}
\noindent\rule{1\linewidth}{1pt}

\begin{table}[t]
\small
\begin{center}
\begin{tabular}{|p{5cm}|p{3cm}|}
\hline
\textbf{Open Question} & \textbf{Methodologies} \\
\hline
How to accelerate block propagation? & PROT, ALG, GAME, SEC, CRYPTO\\
\hline
How to incentivise mining blocks with many valid transactions? & GAME \\
\hline
How to broadcast blocks within a pool? &  ALG, PROT\\
\hline
Efficient network design for mining pools & PROT, ALG \\
\hline
Should pools be allowed? & GAME \\
\hline
How to avoid transaction floods? & SEC, GAME \\
\hline
How to balance DOS prevention and 0-confirmation requirements? & PROT, GAME, ALG\\
\hline
How to model and analyze crypto currency? & PROT, GAME\\
\hline
How to implement a suitable reward system? & GAME \\
\hline
Topology information hiding & SEC, CRYPT, ALG \\
\hline
Giant honest component & ALG\\
\hline
Link failure models & PROT, ALG\\
\hline
How to incentivize nodes to contribute to efficient routing? & PROT, SEC, GAME\\
\hline
Scalable payment routing strategies & PROT, GAME\\
\hline
How to limit congestion? & PROT, ALG\\
\hline
How to balance privacy and efficiency in channels' liquidity observability? & SEC, ALG, CRYPT\\
\hline
How to protect light client from spoofing? & PROT, GAME, SEC\\
\hline
How to protect from isolation and hijacking attacks? & PROT, ALG, GAME\\
\hline
\end{tabular}
\end{center}
\caption{Open network questions and methodologies needed to address them: ALGorithms, network PROTocols, GAME theory, SECurity, CRYPTography.}
\label{table:question}
\vspace{-.5cm}
\end{table}%

\section{Off-chain Payment Channels}\label{sec:payment_channels}

\begin{figure}[t]
	\centering
	\includegraphics[width=0.8\columnwidth]{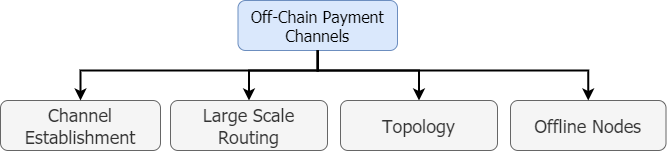}
	\caption{Off-chain Payment Channels Topics}
	\vspace{-.4cm}
\end{figure}

\subsection{What is it about?}

Scaling limitations and transaction latencies have led to a rich corpus 
of work exploring different blockchain scaling solutions: 
including alternative blockchain consensus architectures~\cite{kiayias2017ouroboros,miller2014permacoin,park2018spacemint,kogias2016enhancing,eyal2016Bitcoin}, 
sharding~\cite{luu2016secure,gencer2016service} 
or side-chains (mechanisms that allow digital 
assets from one blockchain to be used in a diffferent ones)~\cite{back2014enabling}, 
to just name a few~\cite{bano2017consensus}. 

Off-chain peer-to-peer networks 
are emerging as a parallel effort to rendering on-chain networks
more scalable.
Off-chain or so-called ``layer-2'' protocols
(built on top of
the layer-1 blockchains) are typically defined
as protocols that do not publish every transaction
on the blockchain immediately (contrary to on-chain
transactions) and entirely rely on the consensus algorithm
of a parent-chain. 
Off-chain protocols rely on channels 
which establish a private peer-to-peer medium, governed
by pre-set rules, e.g., a smart contract, allowing the
involved parties to consent to state updates unanimously by
exchanging authenticated state transitions off-chain.

Channels can come
in different flavors, e.g., 
channels which are formed between $n$
parties,
or
commit-chains, which
 serve a similar purpose as payment channels 
and rely on a central but untrusted intermediary.
One can also distinguish between 
payment channels which are off-chain payment interactions,
and more general 
state channels, which are arbitrary off-chain interactions.
Side-chains~\cite{back2014enabling,bano2017consensus} do not classify as layer-2 solutions due to having
their own consensus algorithm.

Payment channels emerged to support rapid one-way
payments, then transitioned towards bi-directional channel designs
 where both parties can issue and receive payments.
State channels generalize the concept to support the
execution of arbitrary state transitions. 

To give an example, consider 
two companies which transact frequently with each other. 
Rather than settling all their transactions on the blockchain
the companies can each transfer a balance to a unique joint account, which is recorded in the blockchain, such that each party is entitled to receive its
original balance once their joint account is closed. Afterwards, the parties can privately
send money to each other using their cryptographic keys and update the state of their
joint account, without revealing the updated state to the rest of the P2P network. 

Since updates are not always recorded 
in the blockchain, the balance of the state channel 
changes over time, based on the transactions between the companies. When any of the
parties wishes to close the state channel, it can submit the most updated state to the rest
of the network to be recorded in the blockchain, thus closing the state channel and passing
the funds back to the companies based on their most updated balance. State
channels enable the same type of interactions as the Consensus Layer, without recording
them in the blockchain. However they also introduce new complexities. For example, safety
measures are required to prevent a dishonest party from closing of a state channel using
an earlier version of the state channel, thus omitting recent payments.

Popular payment channel networks (PCNs) include Bitcoin Lightning~\cite{lightning}, 
Ethereum Raiden~\cite{raiden}, and XRP Ripple~\cite{fugger2004money}, to name a few.
In all these networks, each node typically represents a user
and each directed, weighted edge represents funds escrowed on a
blockchain; these funds can be transacted only between the endpoints of the edge. 
Many PCNs such as 
Lightning and Raiden use
source routing, in which the source of a payment
specifies the complete route for the payment. If the global
view of all nodes is accurate, source routing is highly effective
because it finds all paths between pairs of nodes. 

 In 
several respects, routing in
PCNs is fairly different from routing in
traditional communication networks: in traditional communication networks, 
routing algorithms typically aim to find short
and low-load paths in a network whose links are subject to
fixed capacity constraints. In a PCN, link
capacities represent payment balances, which can be highly dynamic: 
every transaction changes the payment balance initially
set up for the channel. 

Existing network routing algorithms for data transmission
experience unique challenges when applied to PCNs. 
Node links and bandwidth capacities
in data networks are not considered private information.
In contrast, a PCN routing algorithm
changes the state of the traversed channels to secure
the asset delivery from the sender to the receiver. Depending on the
transaction amount, certain channels may not be suitable to
route a payment, and channel balances are thus an obstacle that
routing algorithms have to account for \cite{saar-economy}. An executed channel
transaction permanently alters the state of all channels along
the path. 

\subsection{State of the Art}

We review state-of-the-art approaches structured around
the different aspects which arise in off-chain
networks: from channel establishment over route discovery to 
supporting scalability.

\noindent \textbf{Routing and Channel Establishment.}
A distinguishing feature of PCNs is that they
also support transactions between participants without direct
channels, using multihop routing. In a nutshell, 
users can efficiently transmit funds from node A to B
by relaying them over a path connecting A to B, as long as each edge
in the path contains enough balance (escrowed funds) to support the
transaction.

However, 
design tradeoffs and security implications of such multi-hop
routing are not well-understood today~\cite{malavolta2019anonymous}.
Scalability is a concern here:
currently, the Lightning network comprises more than $10$k nodes and  $35$k channels,
which are updated and changed frequently. 
We will dedicate a separate subsection to scalability,
and focus on additional aspects in the following.

It has been shown that cost-efficient routing,
aiming to minimize fees, can sometimes be exploited.
Tochner et al.~\cite{tochner2019hijacking} identify 
and analyze a novel DoS attack arising in PCNs,
based on an inherent tradeoff between how efficient (and hence predictable)
versus how secure routes are:
for cost-effectiveness, the routing algorithm should find paths with low transaction fees. The fees of a layer-two transaction
should be lower than the fees for a layer-one transaction.

To provide privacy, routing paths should be found without disclosing
transaction values (i.e. value privacy) and the involved
parties (i.e. sender and receiver privacy).
Tochner et al.'s attack is based on route hijacking 
and exploits the way transactions are routed and executed along the created channels of the network. 
The idea is that an attacker can first create channels that increase
the probability that transactions will route through it. 
Using an amplification attack, 
the attacker can increase the delay of the new channels, delaying
payments for  
the period of the hijack attack.
The authors empirically show that 
with just 5 channels, an attacker can hijack the majority of transactions
($\approx 60\%$, while 30 channels hijack $\approx 90\%$).
The authors also discuss countermeasures,
also showing a simple example where a small change
in the existing weight function of the routing algorithm 
decreases the hijack affect of such attacks.

Similar in spirit is the work by Tang et al.~\cite{tang2019privacy}.
In PCNs, whenever a transaction succeeds, 
edge weights are updated. However, the new channel balances
(i.e., edge weights) are usually not revealed to users directly for privacy reasons. 
This can lead to inefficiencies: when determining a route for
transactions, users first have to guess a path that might be suitable, and then check if it really supports the
transaction. This guess-and-check process dramatically reduces
the success rate of transactions. At the other extreme, knowing
full channel balances can give substantial improvements in success
rate at the expense of privacy. 
To address this problem, Tang et al.~\cite{tang2019privacy}  studied whether a
network can reveal noisy channel balances to trade off privacy for
utility. The authors show that in general, what can be achieved in this context is
fairly limited. They then propose
noise mechanisms 
and find that  it is
not possible to get large gains in utility by giving up a little privacy,
or large gains in privacy by sacrificing a little utility. Hence,
the authors argue that it is optimal to operate either in the low-privacy or low-utility regime.

The routing and hence the performance of the network typically heavily depends on 
the fees. Di Stasi et al.~\cite{8726489} suggested to change
the way how nodes apply fees for forwarding payments,  while trying to keep the network balanced and
improve performance, also using a multipath routing payment scheme, to further reduce the fees paid by users
and keep the network balanced. 
In particular, the authors argue that there are requirements that the fee function should satisfy, and are not currently fulfilled. Di Stasi et al.~also discuss the problem of finding multiple 
paths between a source and a target, to improve transaction routing.

In order to ensure anonymity, onion routing 
is usually used, which however requires the random selection
of nodes in a path.
SilentWhispers~\cite{silentwhispers} and
SpeedyMurmurs~\cite{roos2017settling} formalize and address
concrete notions of privacy in this context.
SpiderNetwork~\cite{sivaraman2018routing} improves the effectiveness of source
routing in a dynamic PCN by favoring routes that minimize the balance difference as well as on-chain
rebalancing, meaning that nodes deposit additional coins to 
improve the balance; their routing relies on a packet-switched
network, that is, instead of routing a complete payment,
the payment is split into constant-size units which
are routed individually, mitigating channel capacity limitations. SpiderNetwork is therefore effective even when balances are constantly changing, at the cost of higher latencies if on-chain rebalancing is used.
	
\noindent \textbf{Routing in Large-Scale Topologies.}
Today's routing algorithms require
every node to know (and maintain) the entire topology,
 in order to be able to compute a route toward the transaction's target. 
This stands in contrast with the objective to support  
lightweight nodes (e.g., wallets) in the network, which should be able to perform transactions as fast as possible, 
requiring minimal disk space and control traffic (e.g., to acknowledge new nodes and channels' updates).
In this sense, the underlying challenges are reminiscent of wireless and adhoc networks~\cite{perkins2001ad}:
nodes are resource-constrained and can decide with whom to 
establish new connections (which comes at a cost).

One of the first and well-known scalable approaches is 
 Flare routing~\cite{prihodko2016flare} with beacon nodes. First, the node proactively maintain a list of channels and beacons in its close neighborhood. The route discovery process first compares the local knowledge with the target, and if a route was not found then it queries the beacons for a route between them. The source node will finally choose a route from the results using heuristic ranks. The incentive of the beacon is to increase the chances that a node will route through it, and it will earn the routing fee.

A major step ahead was then made by the Lightning developers, that relaxed this approach by assuming that nodes can store and maintain the whole network topology and state: Lightning introduces the notion of trampoline nodes~\cite{trampolineblog} to which 
light-weight clients can outsource the route computation. 
In order to find trampoline nodes, light-weight clients can simply use a breadth-first search.
The trampoline nodes know the entire topology (using the current gossiping methodology) and can hence provide routes. To provide incentives Lightning employs fee mechanisms.

A user that wish to find a route can query multiple trampoline nodes, thus the trampoline's incentive is to suggest the best route comparing to his competitors, and maximize the chance that the node will route through the him (and therefore pay the trampoline's fees).

A work by Tochner and Schmid~\cite{saar-economy} analyzes the tradeoff triangle between Confidentiality (using a third party to discover a route), Efficiency (the user's cost to use a route) and Effectiveness (the availability of the route) of the route discovery process. The authors shows analytically that the user can not find a route discovery policy that maximize this tradeoffs, and show empirically that there are topologies in which the tradeoff is bounded.

\noindent \textbf{Topology.}
There are several interesting studies on the topology of transaction networks.
In particular, Rohrer et al.~\cite{rohrer2019discharged} study 
the resilience of the Lightning network to topology-based attacks, 
and in particular, to isolation attacks.
The authors argue that the Lightning network
can be classified as a small-world and scale free network and show that 
in order to perform a resource-limited attack, the attacker should employ a highest ranked minimum cut strategy. 
However, high budgets (around 200BTC) 
are required to give the adversary the power to reliably disturb all payment attempts.
Nisslmueller et al. \cite{icissp20} also showed that
active and passive topology exploration can be exploited 
to attack confidentiality in off-chain networks.

\noindent \textbf{Offline Nodes.}
Typically, intermediaries along a channel route
are required to remain online and explicitly
confirm all mediated transactions. Dziembowski et al.~\cite{dziembowski2017perun,prihodko2016flare} address these shortcomings with the introduction of
virtual channels that support payment and state transitions. All
intermediaries along the route can lock coins for a fixed period
of time and both parties can treat the path as a new virtual
channel connecting them directly. In this manner, A and B
can transact without interacting with intermediaries along the
path, thus reducing the transaction latency. Virtual channels are
limited by the need to recursively set up a new virtual channel
for every intermediary along the path. It is the intermediary's 
responsibility to ensure the channels close appropriately.
In \cite{mccorry2019pisa}, McCorry et al. 
introduce the Pisa protocol which enables parties to delegate to a third party 
to provide security even to parties that may go off-line for an extended period of time. Zeta Avarikioti et al. followed a similar idea in~\cite{avarikioti2019brick}, and suggested to use external parties that both parties of the channel agreed on, and ``approve" any unilaterally action. 

\subsection{Open Problems}

Off-chain networks do not only introduce promising new solutions
but also several new challenges, e.g.,
related to how transactions are routed or information ``gossiped'', or
related to how the topology should be designed.
Indeed, routing and topology become more tightly coupled in offchain
networks, as nodes cannot only strategically choose routes but also
channels: both the establishment as
well as the use of payment channels is an inherently strategic
decision, and subject to complex incentives and the extent to
which a participant thinks she or he can benefit from different
behaviors. A participant may not only try to strategically
maximize her or his profit, but may also be malicious. 

\noindent\rule{1\linewidth}{1pt}
\begin{openq}
\textbf{How should nodes be incentivized 
to contribute to efficient routing?}
Several state-of-the-art works have identified
issues with the current fee-based routing scheme
used, e.g., in Lightning. For example,
by announcing low fees, nodes can launch a
Denial-of-Service attack on transaction routing.
Can we design a pricing scheme which 
on the one hand avoids these issues
and on the other hand still incentivizes
nodes to contribute resources? 
\end{openq}

\begin{openq}
\textbf{How to design payment routing strategies to be more scalable?}
The main question is: Are there better alternatives to source routing?
Several state-of the art approaches such as Flare and Trampoline nodes have been proposed lately, but there is still a long way to go in finding safe and scalable strategies. In particular, routing options today heavily depend on fees, which can cause congestion problems and even be used to devise attacks against the network.
\end{openq}

\begin{openq}
\textbf{How to limit congestion on individual channels?}
This question relates to the previous one. Since today routing depends heavily on fees (which are controlled by the nodes that own the channel), a single node can have a very large effect on the entire network. Is this a case for traffic engineering?
\end{openq}

\begin{openq}
\textbf{Should payments be routable along multiple paths?}
In particular, how can this be achieved in an efficient and secure
manner?
\end{openq}

\begin{openq}
\textbf{Can routing be made aware of the current liquidity balance in channels?} In particular, this raises the question about a possible efficiency-privacy tradeoff. 
\end{openq}

\begin{openq}
\textbf{How can light clients be protected
from spoofing?} E.g., clients may aim to spoof channels, fees on the blockchain, etc.
\end{openq}

\begin{openq}
\textbf{How can we protect against isolation and hijacking attacks?}
We suggest studying whether long liquidity lock attacks can be prevented. In particular, what strategies can the network deploy to prevent the success of such attacks? This question is relevant both in the context of channels, and the gossip network. 
\end{openq}
\noindent\rule{1\linewidth}{1pt}

\section{Conclusion }\label{sec:conclusions}

This paper presented an overview of the 
cryptocurrency networking aspects, with a focus
on open research questions, and towards this goal,
covering also background and state-of-the-art. 
We believe that the networking aspects have
not yet received the attention they deserve,
and hope that our paper can contribute
toward more research in this space.

Table~1 
provides a summary of these questions,
structured around the methodologies needed to address them,
from algorithms, over network protocols, game theory, security,
to cryptography. 

\noindent \textbf{Acknowledgments.}
Research in part supported by the  Vienna Science and Technology Fund (WWTF)
project WHATIF, grant ICT19-045, 2020-2024.


{
\balance
\bibliographystyle{ACM-Reference-Format}
\bibliography{literature}


\begin{thebibliography}{103}


\ifx \showCODEN    \undefined \def \showCODEN     #1{\unskip}     \fi
\ifx \showDOI      \undefined \def \showDOI       #1{#1}\fi
\ifx \showISBNx    \undefined \def \showISBNx     #1{\unskip}     \fi
\ifx \showISBNxiii \undefined \def \showISBNxiii  #1{\unskip}     \fi
\ifx \showISSN     \undefined \def \showISSN      #1{\unskip}     \fi
\ifx \showLCCN     \undefined \def \showLCCN      #1{\unskip}     \fi
\ifx \shownote     \undefined \def \shownote      #1{#1}          \fi
\ifx \showarticletitle \undefined \def \showarticletitle #1{#1}   \fi
\ifx \showURL      \undefined \def \showURL       {\relax}        \fi
\providecommand\bibfield[2]{#2}
\providecommand\bibinfo[2]{#2}
\providecommand\natexlab[1]{#1}
\providecommand\showeprint[2][]{arXiv:#2}

\bibitem[\protect\citeauthoryear{??}{com}{[n.d.]}]%
        {compactblocksrelay}
 \bibinfo{year}{[n.d.]}\natexlab{}.
\newblock \bibinfo{title}{{Bitcoin Improvement Proposals} Compact Block Relay}.
\newblock
  \bibinfo{howpublished}{\url{https://github.com/bitcoin/bips/blob/master/bip-0152.mediawiki}}.
\newblock
\newblock
\shownote{Accessed: 2019-09-23.}


\bibitem[\protect\citeauthoryear{??}{tra}{[n.d.]}]%
        {trampolineblog}
 \bibinfo{year}{[n.d.]}\natexlab{}.
\newblock \bibinfo{title}{Outsourcing Route Computation With Trampoline
  Payments}.
\newblock
  \bibinfo{howpublished}{\url{https://bitcointechweekly.com/front/outsourcing-route-computation-with-trampoline-payments/}}.
\newblock
\newblock
\shownote{Accessed: 2019-04-30.}


\bibitem[\protect\citeauthoryear{Andresen}{Andresen}{2015}]%
        {weakblockthoughts}
\bibfield{author}{\bibinfo{person}{Gavin Andresen}.}
  \bibinfo{year}{2015}\natexlab{}.
\newblock \bibinfo{booktitle}{\emph{Weak Block Thoughts}}.
\newblock
\urldef\tempurl%
\url{https://github.com/ethereum/devp2p/blob/master/caps/eth.md}
\showURL{%
\tempurl}


\bibitem[\protect\citeauthoryear{Apostolaki, Marti, M{\"u}ller, and
  Vanbever}{Apostolaki et~al\mbox{.}}{2018}]%
        {apostolaki2018sabre}
\bibfield{author}{\bibinfo{person}{Maria Apostolaki}, \bibinfo{person}{Gian
  Marti}, \bibinfo{person}{Jan M{\"u}ller}, {and} \bibinfo{person}{Laurent
  Vanbever}.} \bibinfo{year}{2018}\natexlab{}.
\newblock \showarticletitle{SABRE: Protecting Bitcoin against Routing Attacks}.
\newblock \bibinfo{journal}{\emph{arXiv preprint arXiv:1808.06254}}
  (\bibinfo{year}{2018}).
\newblock


\bibitem[\protect\citeauthoryear{Apostolaki, Zohar, and Vanbever}{Apostolaki
  et~al\mbox{.}}{2017}]%
        {apostolaki2017hijacking}
\bibfield{author}{\bibinfo{person}{Maria Apostolaki}, \bibinfo{person}{Aviv
  Zohar}, {and} \bibinfo{person}{Laurent Vanbever}.}
  \bibinfo{year}{2017}\natexlab{}.
\newblock \showarticletitle{Hijacking bitcoin: Routing attacks on
  cryptocurrencies}. In \bibinfo{booktitle}{\emph{2017 IEEE Symposium on
  Security and Privacy (SP)}}. IEEE, \bibinfo{pages}{375--392}.
\newblock


\bibitem[\protect\citeauthoryear{Avarikioti, Kogias, and
  Wattenhofer}{Avarikioti et~al\mbox{.}}{2019}]%
        {avarikioti2019brick}
\bibfield{author}{\bibinfo{person}{Georgia Avarikioti},
  \bibinfo{person}{Eleftherios~Kokoris Kogias}, {and} \bibinfo{person}{Roger
  Wattenhofer}.} \bibinfo{year}{2019}\natexlab{}.
\newblock \showarticletitle{Brick: Asynchronous state channels}.
\newblock \bibinfo{journal}{\emph{arXiv preprint arXiv:1905.11360}}
  (\bibinfo{year}{2019}).
\newblock


\bibitem[\protect\citeauthoryear{Babaioff, Dobzinski, Oren, and Zohar}{Babaioff
  et~al\mbox{.}}{2012}]%
        {babaioff2012Bitcoin}
\bibfield{author}{\bibinfo{person}{Moshe Babaioff}, \bibinfo{person}{Shahar
  Dobzinski}, \bibinfo{person}{Sigal Oren}, {and} \bibinfo{person}{Aviv
  Zohar}.} \bibinfo{year}{2012}\natexlab{}.
\newblock \showarticletitle{On bitcoin and red balloons}. In
  \bibinfo{booktitle}{\emph{Proceedings of the 13th ACM conference on
  electronic commerce}}. \bibinfo{pages}{56--73}.
\newblock


\bibitem[\protect\citeauthoryear{Back, Corallo, Dashjr, Friedenbach, Maxwell,
  Miller, Poelstra, Tim{\'o}n, and Wuille}{Back et~al\mbox{.}}{2014}]%
        {back2014enabling}
\bibfield{author}{\bibinfo{person}{Adam Back}, \bibinfo{person}{Matt Corallo},
  \bibinfo{person}{Luke Dashjr}, \bibinfo{person}{Mark Friedenbach},
  \bibinfo{person}{Gregory Maxwell}, \bibinfo{person}{Andrew Miller},
  \bibinfo{person}{Andrew Poelstra}, \bibinfo{person}{Jorge Tim{\'o}n}, {and}
  \bibinfo{person}{Pieter Wuille}.} \bibinfo{year}{2014}\natexlab{}.
\newblock \showarticletitle{Enabling blockchain innovations with pegged
  sidechains}.
\newblock \bibinfo{journal}{\emph{URL: http://www. opensciencereview.
  com/papers/123/enablingblockchain-innovations-with-pegged-sidechains}}
  \bibinfo{volume}{72} (\bibinfo{year}{2014}).
\newblock


\bibitem[\protect\citeauthoryear{Bano, Sonnino, Al-Bassam, Azouvi, McCorry,
  Meiklejohn, and Danezis}{Bano et~al\mbox{.}}{2017}]%
        {bano2017consensus}
\bibfield{author}{\bibinfo{person}{Shehar Bano}, \bibinfo{person}{Alberto
  Sonnino}, \bibinfo{person}{Mustafa Al-Bassam}, \bibinfo{person}{Sarah
  Azouvi}, \bibinfo{person}{Patrick McCorry}, \bibinfo{person}{Sarah
  Meiklejohn}, {and} \bibinfo{person}{George Danezis}.}
  \bibinfo{year}{2017}\natexlab{}.
\newblock \showarticletitle{Consensus in the age of blockchains}.
\newblock \bibinfo{journal}{\emph{arXiv preprint arXiv:1711.03936}}
  (\bibinfo{year}{2017}).
\newblock


\bibitem[\protect\citeauthoryear{Basu, Eyal, and Sirer}{Basu
  et~al\mbox{.}}{2016}]%
        {falcon}
\bibfield{author}{\bibinfo{person}{Soumaya Basu}, \bibinfo{person}{Ittay Eyal},
  {and} \bibinfo{person}{Emin~Gun Sirer}.} \bibinfo{year}{2016}\natexlab{}.
\newblock \bibinfo{booktitle}{\emph{Falcon: Relay Network for Bitcoin Blocks}}.
\newblock
\urldef\tempurl%
\url{https://www.falcon-net.org/}
\showURL{%
\tempurl}


\bibitem[\protect\citeauthoryear{Ben~Mariem, Casas, and Donnet}{Ben~Mariem
  et~al\mbox{.}}{2018}]%
        {ben2018vivisecting}
\bibfield{author}{\bibinfo{person}{Sami Ben~Mariem}, \bibinfo{person}{Pedro
  Casas}, {and} \bibinfo{person}{Beno{\^\i}t Donnet}.}
  \bibinfo{year}{2018}\natexlab{}.
\newblock \showarticletitle{Vivisecting Blockchain P2P Networks: Unveiling the
  Bitcoin IP Network}. In \bibinfo{booktitle}{\emph{ACM CoNEXT Student
  Workshop}}.
\newblock


\bibitem[\protect\citeauthoryear{Biryukov, Khovratovich, and
  Pustogarov}{Biryukov et~al\mbox{.}}{2014}]%
        {biryukov2014deanonymisation}
\bibfield{author}{\bibinfo{person}{Alex Biryukov}, \bibinfo{person}{Dmitry
  Khovratovich}, {and} \bibinfo{person}{Ivan Pustogarov}.}
  \bibinfo{year}{2014}\natexlab{}.
\newblock \showarticletitle{Deanonymisation of clients in Bitcoin P2P network}.
  In \bibinfo{booktitle}{\emph{Proceedings of the 2014 ACM SIGSAC Conference on
  Computer and Communications Security}}. \bibinfo{pages}{15--29}.
\newblock


\bibitem[\protect\citeauthoryear{Bitcoin}{Bitcoin}{2019}]%
        {Bitcoin-p2p-guide}
\bibfield{author}{\bibinfo{person}{Bitcoin}.} \bibinfo{year}{2019}\natexlab{}.
\newblock \bibinfo{booktitle}{\emph{P2P Guide}}.
\newblock
\urldef\tempurl%
\url{https://bitcoin.org/en/p2p-network-guide}
\showURL{%
\tempurl}


\bibitem[\protect\citeauthoryear{BlueMatt}{BlueMatt}{2016}]%
        {fibre}
\bibfield{author}{\bibinfo{person}{BlueMatt}.} \bibinfo{year}{2016}\natexlab{}.
\newblock \bibinfo{title}{FIBRE}.
\newblock
\newblock
\urldef\tempurl%
\url{http://bitcoinfibre.org/public-network.html}
\showURL{%
\tempurl}


\bibitem[\protect\citeauthoryear{Bojja~Venkatakrishnan, Fanti, and
  Viswanath}{Bojja~Venkatakrishnan et~al\mbox{.}}{2017}]%
        {bojja2017dandelion}
\bibfield{author}{\bibinfo{person}{Shaileshh Bojja~Venkatakrishnan},
  \bibinfo{person}{Giulia Fanti}, {and} \bibinfo{person}{Pramod Viswanath}.}
  \bibinfo{year}{2017}\natexlab{}.
\newblock \showarticletitle{Dandelion: Redesigning the bitcoin network for
  anonymity}.
\newblock \bibinfo{journal}{\emph{Proceedings of the ACM on Measurement and
  Analysis of Computing Systems}} \bibinfo{volume}{1}, \bibinfo{number}{1}
  (\bibinfo{year}{2017}), \bibinfo{pages}{22}.
\newblock


\bibitem[\protect\citeauthoryear{Bonneau, Miller, Clark, Narayanan, Kroll, and
  Felten}{Bonneau et~al\mbox{.}}{2015}]%
        {bonneau2015sok}
\bibfield{author}{\bibinfo{person}{Joseph Bonneau}, \bibinfo{person}{Andrew
  Miller}, \bibinfo{person}{Jeremy Clark}, \bibinfo{person}{Arvind Narayanan},
  \bibinfo{person}{Joshua~A Kroll}, {and} \bibinfo{person}{Edward~W Felten}.}
  \bibinfo{year}{2015}\natexlab{}.
\newblock \showarticletitle{Sok: Research perspectives and challenges for
  bitcoin and cryptocurrencies}. In \bibinfo{booktitle}{\emph{Proc. IEEE
  Symposium on Security and Privacy (SP)}}. IEEE, \bibinfo{pages}{104--121}.
\newblock


\bibitem[\protect\citeauthoryear{Bortnikov, Gurevich, Keidar, Kliot, and
  Shraer}{Bortnikov et~al\mbox{.}}{2009}]%
        {bortnikov2009brahms}
\bibfield{author}{\bibinfo{person}{Edward Bortnikov}, \bibinfo{person}{Maxim
  Gurevich}, \bibinfo{person}{Idit Keidar}, \bibinfo{person}{Gabriel Kliot},
  {and} \bibinfo{person}{Alexander Shraer}.} \bibinfo{year}{2009}\natexlab{}.
\newblock \showarticletitle{Brahms: Byzantine resilient random membership
  sampling}.
\newblock \bibinfo{journal}{\emph{Computer Networks}} \bibinfo{volume}{53},
  \bibinfo{number}{13} (\bibinfo{year}{2009}), \bibinfo{pages}{2340--2359}.
\newblock


\bibitem[\protect\citeauthoryear{Cao, Yu, Decouchant, Luo, and
  Ver{\'\i}ssimo}{Cao et~al\mbox{.}}{2019}]%
        {cao2019exploring}
\bibfield{author}{\bibinfo{person}{Tong Cao}, \bibinfo{person}{Jiangshan Yu},
  \bibinfo{person}{J{\'e}r{\'e}mie Decouchant}, \bibinfo{person}{Xiapu Luo},
  {and} \bibinfo{person}{Paulo Ver{\'\i}ssimo}.}
  \bibinfo{year}{2019}\natexlab{}.
\newblock \showarticletitle{Exploring the Monero Peer-to-Peer Network.}
\newblock \bibinfo{journal}{\emph{IACR Cryptology ePrint Archive}}
  \bibinfo{volume}{2019} (\bibinfo{year}{2019}), \bibinfo{pages}{411}.
\newblock


\bibitem[\protect\citeauthoryear{Cardano}{Cardano}{2019}]%
        {cardano-topology}
\bibfield{author}{\bibinfo{person}{Cardano}.} \bibinfo{year}{2019}\natexlab{}.
\newblock \bibinfo{booktitle}{\emph{P2P Topology}}.
\newblock
\urldef\tempurl%
\url{https://cardanodocs.com/cardano/topology/}
\showURL{%
\tempurl}


\bibitem[\protect\citeauthoryear{Chawla, Behrens, Tapp, Boscovic, and
  Candan}{Chawla et~al\mbox{.}}{2019}]%
        {chawla2019velocity}
\bibfield{author}{\bibinfo{person}{Nakul Chawla}, \bibinfo{person}{Hans~Walter
  Behrens}, \bibinfo{person}{Darren Tapp}, \bibinfo{person}{Dragan Boscovic},
  {and} \bibinfo{person}{K~Sel{\c{c}}uk Candan}.}
  \bibinfo{year}{2019}\natexlab{}.
\newblock \showarticletitle{Velocity: Scalability improvements in block
  propagation through rateless erasure coding}. In
  \bibinfo{booktitle}{\emph{2019 IEEE International Conference on Blockchain
  and Cryptocurrency (ICBC)}}. IEEE, \bibinfo{pages}{447--454}.
\newblock


\bibitem[\protect\citeauthoryear{Decker and Wattenhofer}{Decker and
  Wattenhofer}{2013}]%
        {decker2013information}
\bibfield{author}{\bibinfo{person}{Christian Decker} {and}
  \bibinfo{person}{Roger Wattenhofer}.} \bibinfo{year}{2013}\natexlab{}.
\newblock \showarticletitle{Information propagation in the bitcoin network}. In
  \bibinfo{booktitle}{\emph{Proc. IEEE P2P Conference}}. IEEE,
  \bibinfo{pages}{1--10}.
\newblock


\bibitem[\protect\citeauthoryear{Delgado-Segura, Bakshi, P{\'e}rez-Sol{\`a},
  Litton, Pachulski, Miller, and Bhattacharjee}{Delgado-Segura
  et~al\mbox{.}}{2018a}]%
        {delgado2018txprobe}
\bibfield{author}{\bibinfo{person}{Sergi Delgado-Segura},
  \bibinfo{person}{Surya Bakshi}, \bibinfo{person}{Cristina
  P{\'e}rez-Sol{\`a}}, \bibinfo{person}{James Litton}, \bibinfo{person}{Andrew
  Pachulski}, \bibinfo{person}{Andrew Miller}, {and} \bibinfo{person}{Bobby
  Bhattacharjee}.} \bibinfo{year}{2018}\natexlab{a}.
\newblock \showarticletitle{TxProbe: Discovering Bitcoin's Network Topology
  Using Orphan Transactions}.
\newblock \bibinfo{journal}{\emph{arXiv preprint arXiv:1812.00942}}
  (\bibinfo{year}{2018}).
\newblock


\bibitem[\protect\citeauthoryear{Delgado-Segura, P{\'e}rez-Sol{\`a},
  Herrera-Joancomart{\'\i}, Navarro-Arribas, and Borrell}{Delgado-Segura
  et~al\mbox{.}}{2018b}]%
        {delgado2018cryptocurrency}
\bibfield{author}{\bibinfo{person}{Sergi Delgado-Segura},
  \bibinfo{person}{Cristina P{\'e}rez-Sol{\`a}}, \bibinfo{person}{Jordi
  Herrera-Joancomart{\'\i}}, \bibinfo{person}{Guillermo Navarro-Arribas}, {and}
  \bibinfo{person}{Joan Borrell}.} \bibinfo{year}{2018}\natexlab{b}.
\newblock \showarticletitle{Cryptocurrency networks: A new p2p paradigm}.
\newblock \bibinfo{journal}{\emph{Mobile Information Systems}}
  \bibinfo{volume}{2018} (\bibinfo{year}{2018}).
\newblock


\bibitem[\protect\citeauthoryear{{Di Stasi}, {Avallone}, {Canonico}, and
  {Ventre}}{{Di Stasi} et~al\mbox{.}}{2018}]%
        {8726489}
\bibfield{author}{\bibinfo{person}{G. {Di Stasi}}, \bibinfo{person}{S.
  {Avallone}}, \bibinfo{person}{R. {Canonico}}, {and} \bibinfo{person}{G.
  {Ventre}}.} \bibinfo{year}{2018}\natexlab{}.
\newblock \showarticletitle{Routing Payments on the Lightning Network}. In
  \bibinfo{booktitle}{\emph{Proc. IEEE Blockchain}}.
  \bibinfo{pages}{1161--1170}.
\newblock


\bibitem[\protect\citeauthoryear{Donet, P{\'e}rez-Sola, and
  Herrera-Joancomart{\'\i}}{Donet et~al\mbox{.}}{2014}]%
        {donet2014Bitcoin}
\bibfield{author}{\bibinfo{person}{Joan~Antoni Donet},
  \bibinfo{person}{Cristina P{\'e}rez-Sola}, {and} \bibinfo{person}{Jordi
  Herrera-Joancomart{\'\i}}.} \bibinfo{year}{2014}\natexlab{}.
\newblock \showarticletitle{The bitcoin P2P network}. In
  \bibinfo{booktitle}{\emph{International Conference on Financial Cryptography
  and Data Security}}. Springer, \bibinfo{pages}{87--102}.
\newblock


\bibitem[\protect\citeauthoryear{Drucker and Fleischer}{Drucker and
  Fleischer}{2012}]%
        {sybil-proof-marketing}
\bibfield{author}{\bibinfo{person}{Fabio~A Drucker} {and}
  \bibinfo{person}{Lisa~K Fleischer}.} \bibinfo{year}{2012}\natexlab{}.
\newblock \showarticletitle{Simpler sybil-proof mechanisms for multi-level
  marketing}. In \bibinfo{booktitle}{\emph{Proceedings of the 13th ACM
  conference on Electronic commerce}}. \bibinfo{pages}{441--458}.
\newblock


\bibitem[\protect\citeauthoryear{Dyagilev, Mannor, and Yom-Tov}{Dyagilev
  et~al\mbox{.}}{2010}]%
        {rapid-information-propagation}
\bibfield{author}{\bibinfo{person}{Kirill Dyagilev}, \bibinfo{person}{Shie
  Mannor}, {and} \bibinfo{person}{Elad Yom-Tov}.}
  \bibinfo{year}{2010}\natexlab{}.
\newblock \showarticletitle{Generative models for rapid information
  propagation}. In \bibinfo{booktitle}{\emph{Proceedings of the First Workshop
  on Social Media Analytics}}. \bibinfo{pages}{35--43}.
\newblock


\bibitem[\protect\citeauthoryear{Dziembowski, Eckey, Faust, and
  Malinowski}{Dziembowski et~al\mbox{.}}{2017}]%
        {dziembowski2017perun}
\bibfield{author}{\bibinfo{person}{Stefan Dziembowski}, \bibinfo{person}{Lisa
  Eckey}, \bibinfo{person}{Sebastian Faust}, {and} \bibinfo{person}{Daniel
  Malinowski}.} \bibinfo{year}{2017}\natexlab{}.
\newblock \showarticletitle{PERUN: Virtual Payment Channels over Cryptographic
  Currencies.}
\newblock \bibinfo{journal}{\emph{IACR Cryptology ePrint Archive}}
  \bibinfo{volume}{2017} (\bibinfo{year}{2017}), \bibinfo{pages}{635}.
\newblock


\bibitem[\protect\citeauthoryear{Ekparinya, Gramoli, and Jourjon}{Ekparinya
  et~al\mbox{.}}{2018}]%
        {ekparinya2018impact}
\bibfield{author}{\bibinfo{person}{Parinya Ekparinya}, \bibinfo{person}{Vincent
  Gramoli}, {and} \bibinfo{person}{Guillaume Jourjon}.}
  \bibinfo{year}{2018}\natexlab{}.
\newblock \showarticletitle{Impact of man-in-the-middle attacks on ethereum}.
  In \bibinfo{booktitle}{\emph{2018 IEEE 37th Symposium on Reliable Distributed
  Systems (SRDS)}}. IEEE, \bibinfo{pages}{11--20}.
\newblock


\bibitem[\protect\citeauthoryear{Ersoy, Ren, Erkin, and Lagendijk}{Ersoy
  et~al\mbox{.}}{2018}]%
        {ersoy2018transaction}
\bibfield{author}{\bibinfo{person}{O{\u{g}}uzhan Ersoy},
  \bibinfo{person}{Zhijie Ren}, \bibinfo{person}{Zekeriya Erkin}, {and}
  \bibinfo{person}{Reginald~L Lagendijk}.} \bibinfo{year}{2018}\natexlab{}.
\newblock \showarticletitle{Transaction propagation on permissionless
  blockchains: incentive and routing mechanisms}. In
  \bibinfo{booktitle}{\emph{2018 Crypto Valley Conference on Blockchain
  Technology (CVCBT)}}. IEEE, \bibinfo{pages}{20--30}.
\newblock


\bibitem[\protect\citeauthoryear{Ethereum}{Ethereum}{2020}]%
        {wireprotocolehthereum}
\bibfield{author}{\bibinfo{person}{Ethereum}.} \bibinfo{year}{2020}\natexlab{}.
\newblock \bibinfo{booktitle}{\emph{Ethereum Wire Protocol}}.
\newblock
\urldef\tempurl%
\url{https://github.com/ethereum/devp2p/blob/master/caps/eth.md}
\showURL{%
\tempurl}


\bibitem[\protect\citeauthoryear{Eyal, Gencer, Sirer, and Van~Renesse}{Eyal
  et~al\mbox{.}}{2016}]%
        {eyal2016Bitcoin}
\bibfield{author}{\bibinfo{person}{Ittay Eyal}, \bibinfo{person}{Adem~Efe
  Gencer}, \bibinfo{person}{Emin~G{\"u}n Sirer}, {and} \bibinfo{person}{Robbert
  Van~Renesse}.} \bibinfo{year}{2016}\natexlab{}.
\newblock \showarticletitle{Bitcoin-ng: A scalable blockchain protocol}. In
  \bibinfo{booktitle}{\emph{13th USENIX Symposium on Networked Systems Design
  and Implementation (NSDI)}}. \bibinfo{pages}{45--59}.
\newblock


\bibitem[\protect\citeauthoryear{Eyal and Sirer}{Eyal and Sirer}{2014}]%
        {eyal2014majority}
\bibfield{author}{\bibinfo{person}{Ittay Eyal} {and}
  \bibinfo{person}{Emin~G{\"u}n Sirer}.} \bibinfo{year}{2014}\natexlab{}.
\newblock \showarticletitle{Majority is not enough: Bitcoin mining is
  vulnerable}. In \bibinfo{booktitle}{\emph{International conference on
  financial cryptography and data security}}.
\newblock


\bibitem[\protect\citeauthoryear{Eyal and Sirer}{Eyal and Sirer}{2018}]%
        {eyal2018majority}
\bibfield{author}{\bibinfo{person}{Ittay Eyal} {and}
  \bibinfo{person}{Emin~G{\"u}n Sirer}.} \bibinfo{year}{2018}\natexlab{}.
\newblock \showarticletitle{Majority is not enough: Bitcoin mining is
  vulnerable}.
\newblock \bibinfo{journal}{\emph{Commun. ACM}} \bibinfo{volume}{61},
  \bibinfo{number}{7} (\bibinfo{year}{2018}), \bibinfo{pages}{95--102}.
\newblock


\bibitem[\protect\citeauthoryear{Fanti, Venkatakrishnan, Bakshi, Denby,
  Bhargava, Miller, and Viswanath}{Fanti et~al\mbox{.}}{2018}]%
        {fanti2018dandelion++}
\bibfield{author}{\bibinfo{person}{Giulia Fanti},
  \bibinfo{person}{Shaileshh~Bojja Venkatakrishnan}, \bibinfo{person}{Surya
  Bakshi}, \bibinfo{person}{Bradley Denby}, \bibinfo{person}{Shruti Bhargava},
  \bibinfo{person}{Andrew Miller}, {and} \bibinfo{person}{Pramod Viswanath}.}
  \bibinfo{year}{2018}\natexlab{}.
\newblock \showarticletitle{Dandelion++: Lightweight cryptocurrency networking
  with formal anonymity guarantees}.
\newblock \bibinfo{journal}{\emph{Proceedings of the ACM on Measurement and
  Analysis of Computing Systems}} \bibinfo{volume}{2}, \bibinfo{number}{2}
  (\bibinfo{year}{2018}), \bibinfo{pages}{29}.
\newblock


\bibitem[\protect\citeauthoryear{Feld, Sch{\"o}nfeld, and Werner}{Feld
  et~al\mbox{.}}{2014}]%
        {feld2014analyzing}
\bibfield{author}{\bibinfo{person}{Sebastian Feld}, \bibinfo{person}{Mirco
  Sch{\"o}nfeld}, {and} \bibinfo{person}{Martin Werner}.}
  \bibinfo{year}{2014}\natexlab{}.
\newblock \showarticletitle{Analyzing the Deployment of Bitcoin's P2P Network
  under an AS-level Perspective}.
\newblock \bibinfo{journal}{\emph{Procedia Computer Science}}
  \bibinfo{volume}{32} (\bibinfo{year}{2014}), \bibinfo{pages}{1121--1126}.
\newblock


\bibitem[\protect\citeauthoryear{Fugger}{Fugger}{2004}]%
        {fugger2004money}
\bibfield{author}{\bibinfo{person}{Ryan Fugger}.}
  \bibinfo{year}{2004}\natexlab{}.
\newblock \showarticletitle{Money as IOUs in social trust networks \& a
  proposal for a decentralized currency network protocol}.
\newblock \bibinfo{journal}{\emph{Hypertext document. Available electronically
  at http://ripple. sourceforge. net}}  \bibinfo{volume}{106}
  (\bibinfo{year}{2004}).
\newblock


\bibitem[\protect\citeauthoryear{Gencer, Basu, Eyal, Van~Renesse, and
  Sirer}{Gencer et~al\mbox{.}}{2018}]%
        {gencer2018decentralization}
\bibfield{author}{\bibinfo{person}{Adem~Efe Gencer}, \bibinfo{person}{Soumya
  Basu}, \bibinfo{person}{Ittay Eyal}, \bibinfo{person}{Robbert Van~Renesse},
  {and} \bibinfo{person}{Emin~G{\"u}n Sirer}.} \bibinfo{year}{2018}\natexlab{}.
\newblock \showarticletitle{Decentralization in bitcoin and ethereum networks}.
\newblock \bibinfo{journal}{\emph{arXiv preprint arXiv:1801.03998}}
  (\bibinfo{year}{2018}).
\newblock


\bibitem[\protect\citeauthoryear{Gencer, van Renesse, and Sirer}{Gencer
  et~al\mbox{.}}{2016}]%
        {gencer2016service}
\bibfield{author}{\bibinfo{person}{Adem~Efe Gencer}, \bibinfo{person}{Robbert
  van Renesse}, {and} \bibinfo{person}{Emin~G{\"u}n Sirer}.}
  \bibinfo{year}{2016}\natexlab{}.
\newblock \showarticletitle{Service-oriented sharding with aspen}.
\newblock \bibinfo{journal}{\emph{arXiv preprint arXiv:1611.06816}}
  (\bibinfo{year}{2016}).
\newblock


\bibitem[\protect\citeauthoryear{Gervais, Capkun, Karame, and Gruber}{Gervais
  et~al\mbox{.}}{2014}]%
        {gervais2014privacy}
\bibfield{author}{\bibinfo{person}{Arthur Gervais}, \bibinfo{person}{Srdjan
  Capkun}, \bibinfo{person}{Ghassan~O Karame}, {and} \bibinfo{person}{Damian
  Gruber}.} \bibinfo{year}{2014}\natexlab{}.
\newblock \showarticletitle{On the privacy provisions of bloom filters in
  lightweight bitcoin clients}. In \bibinfo{booktitle}{\emph{Proceedings of the
  30th Annual Computer Security Applications Conference}}.
  \bibinfo{pages}{326--335}.
\newblock


\bibitem[\protect\citeauthoryear{Gervais, Karame, W{\"u}st, Glykantzis,
  Ritzdorf, and Capkun}{Gervais et~al\mbox{.}}{2016}]%
        {gervais2016security}
\bibfield{author}{\bibinfo{person}{Arthur Gervais}, \bibinfo{person}{Ghassan~O
  Karame}, \bibinfo{person}{Karl W{\"u}st}, \bibinfo{person}{Vasileios
  Glykantzis}, \bibinfo{person}{Hubert Ritzdorf}, {and} \bibinfo{person}{Srdjan
  Capkun}.} \bibinfo{year}{2016}\natexlab{}.
\newblock \showarticletitle{On the security and performance of proof of work
  blockchains}. In \bibinfo{booktitle}{\emph{Proc. of the 2016 ACM SIGSAC
  conference on computer and communications security}}.
\newblock


\bibitem[\protect\citeauthoryear{Goldschlag, Reed, and Syverson}{Goldschlag
  et~al\mbox{.}}{1999}]%
        {goldschlag1999onion}
\bibfield{author}{\bibinfo{person}{David Goldschlag}, \bibinfo{person}{Michael
  Reed}, {and} \bibinfo{person}{Paul Syverson}.}
  \bibinfo{year}{1999}\natexlab{}.
\newblock \bibinfo{booktitle}{\emph{Onion routing for anonymous and private
  internet connections}}.
\newblock \bibinfo{type}{{T}echnical {R}eport}. \bibinfo{institution}{NAVAL
  RESEARCH LAB WASHINGTON DC CENTER FOR HIGH ASSURANCE COMPUTING SYSTEMS~…}.
\newblock


\bibitem[\protect\citeauthoryear{Goodrich and Mitzenmacher}{Goodrich and
  Mitzenmacher}{2011}]%
        {goodrich2011invertible}
\bibfield{author}{\bibinfo{person}{Michael~T Goodrich} {and}
  \bibinfo{person}{Michael Mitzenmacher}.} \bibinfo{year}{2011}\natexlab{}.
\newblock \showarticletitle{Invertible bloom lookup tables}. In
  \bibinfo{booktitle}{\emph{Allerton}}. IEEE.
\newblock


\bibitem[\protect\citeauthoryear{Grundmann, Neudecker, and
  Hartenstein}{Grundmann et~al\mbox{.}}{2018}]%
        {grundmann2018exploiting}
\bibfield{author}{\bibinfo{person}{Matthias Grundmann}, \bibinfo{person}{Till
  Neudecker}, {and} \bibinfo{person}{Hannes Hartenstein}.}
  \bibinfo{year}{2018}\natexlab{}.
\newblock \showarticletitle{Exploiting transaction accumulation and double
  spends for topology inference in bitcoin}. In
  \bibinfo{booktitle}{\emph{International Conference on Financial Cryptography
  and Data Security}}. Springer, \bibinfo{pages}{113--126}.
\newblock


\bibitem[\protect\citeauthoryear{Gudgeon, Moreno-Sanchez, Roos, McCorry, and
  Gervais}{Gudgeon et~al\mbox{.}}{2019}]%
        {gudgeon2019sok}
\bibfield{author}{\bibinfo{person}{Lewis Gudgeon}, \bibinfo{person}{Pedro
  Moreno-Sanchez}, \bibinfo{person}{Stefanie Roos}, \bibinfo{person}{Patrick
  McCorry}, {and} \bibinfo{person}{Arthur Gervais}.}
  \bibinfo{year}{2019}\natexlab{}.
\newblock \showarticletitle{SoK: Off The Chain Transactions.}
\newblock \bibinfo{journal}{\emph{IACR Cryptology ePrint Archive}}
  \bibinfo{volume}{2019} (\bibinfo{year}{2019}), \bibinfo{pages}{360}.
\newblock


\bibitem[\protect\citeauthoryear{Guerraoui, Huc, and Kermarrec}{Guerraoui
  et~al\mbox{.}}{2013}]%
        {guerraoui2013highly}
\bibfield{author}{\bibinfo{person}{Rachid Guerraoui}, \bibinfo{person}{Florian
  Huc}, {and} \bibinfo{person}{Anne-Marie Kermarrec}.}
  \bibinfo{year}{2013}\natexlab{}.
\newblock \showarticletitle{Highly dynamic distributed computing with byzantine
  failures}. In \bibinfo{booktitle}{\emph{Proceedings of the 2013 ACM symposium
  on Principles of distributed computing}}. \bibinfo{pages}{176--183}.
\newblock


\bibitem[\protect\citeauthoryear{Haeupler}{Haeupler}{2016}]%
        {haeupler2016analyzing}
\bibfield{author}{\bibinfo{person}{Bernhard Haeupler}.}
  \bibinfo{year}{2016}\natexlab{}.
\newblock \showarticletitle{Analyzing network coding (gossip) made easy}.
\newblock \bibinfo{journal}{\emph{Journal of the ACM (JACM)}}
  \bibinfo{volume}{63}, \bibinfo{number}{3} (\bibinfo{year}{2016}),
  \bibinfo{pages}{1--22}.
\newblock


\bibitem[\protect\citeauthoryear{Heilman, Kendler, Zohar, and Goldberg}{Heilman
  et~al\mbox{.}}{2015}]%
        {heilman2015eclipse}
\bibfield{author}{\bibinfo{person}{Ethan Heilman}, \bibinfo{person}{Alison
  Kendler}, \bibinfo{person}{Aviv Zohar}, {and} \bibinfo{person}{Sharon
  Goldberg}.} \bibinfo{year}{2015}\natexlab{}.
\newblock \showarticletitle{Eclipse attacks on bitcoin’s peer-to-peer
  network}. In \bibinfo{booktitle}{\emph{24th $\{$USENIX$\}$ Security
  Symposium}}.
\newblock


\bibitem[\protect\citeauthoryear{Imtiaz, Starobinski, Trachtenberg, and
  Younis}{Imtiaz et~al\mbox{.}}{2019}]%
        {imtiaz2019churn}
\bibfield{author}{\bibinfo{person}{Muhammad~Anas Imtiaz},
  \bibinfo{person}{David Starobinski}, \bibinfo{person}{Ari Trachtenberg},
  {and} \bibinfo{person}{Nabeel Younis}.} \bibinfo{year}{2019}\natexlab{}.
\newblock \showarticletitle{Churn in the Bitcoin Network: Characterization and
  Impact}. In \bibinfo{booktitle}{\emph{2019 IEEE International Conference on
  Blockchain and Cryptocurrency (ICBC)}}. IEEE, \bibinfo{pages}{431--439}.
\newblock


\bibitem[\protect\citeauthoryear{Inc.}{Inc.}{2015}]%
        {stratumBitcoinit}
\bibfield{author}{\bibinfo{person}{Bitcoin Inc.}}
  \bibinfo{year}{2015}\natexlab{}.
\newblock \bibinfo{booktitle}{\emph{Stratum Mining Protocol}}.
\newblock
\urldef\tempurl%
\url{https://en.bitcoin.it/wiki/Stratum_mining_protocol}
\showURL{%
\tempurl}


\bibitem[\protect\citeauthoryear{Inc.}{Inc.}{2018}]%
        {networkBitcoinit}
\bibfield{author}{\bibinfo{person}{Bitcoin Inc.}}
  \bibinfo{year}{2018}\natexlab{}.
\newblock \bibinfo{booktitle}{\emph{Network}}.
\newblock
\urldef\tempurl%
\url{https://en.bitcoin.it/wiki/Network}
\showURL{%
\tempurl}


\bibitem[\protect\citeauthoryear{Jaiyeola, Patron, Saia, Young, and
  Zhou}{Jaiyeola et~al\mbox{.}}{2018}]%
        {jaiyeola2018tiny}
\bibfield{author}{\bibinfo{person}{Mercy~O Jaiyeola}, \bibinfo{person}{Kyle
  Patron}, \bibinfo{person}{Jared Saia}, \bibinfo{person}{Maxwell Young}, {and}
  \bibinfo{person}{Qian~M Zhou}.} \bibinfo{year}{2018}\natexlab{}.
\newblock \showarticletitle{Tiny Groups Tackle Byzantine Adversaries}. In
  \bibinfo{booktitle}{\emph{2018 IEEE International Parallel and Distributed
  Processing Symposium (IPDPS)}}. IEEE, \bibinfo{pages}{1030--1039}.
\newblock


\bibitem[\protect\citeauthoryear{Johnson, Laszka, Grossklags, Vasek, and
  Moore}{Johnson et~al\mbox{.}}{2014}]%
        {johnson2014game}
\bibfield{author}{\bibinfo{person}{Benjamin Johnson}, \bibinfo{person}{Aron
  Laszka}, \bibinfo{person}{Jens Grossklags}, \bibinfo{person}{Marie Vasek},
  {and} \bibinfo{person}{Tyler Moore}.} \bibinfo{year}{2014}\natexlab{}.
\newblock \showarticletitle{Game-theoretic analysis of DDoS attacks against
  Bitcoin mining pools}. In \bibinfo{booktitle}{\emph{International Conference
  on Financial Cryptography and Data Security}}. Springer,
  \bibinfo{pages}{72--86}.
\newblock


\bibitem[\protect\citeauthoryear{Karame, Androulaki, and Capkun}{Karame
  et~al\mbox{.}}{2012}]%
        {Karame:2012:DFP:2382196.2382292}
\bibfield{author}{\bibinfo{person}{Ghassan~O. Karame}, \bibinfo{person}{Elli
  Androulaki}, {and} \bibinfo{person}{Srdjan Capkun}.}
  \bibinfo{year}{2012}\natexlab{}.
\newblock \showarticletitle{Double-spending Fast Payments in Bitcoin}. In
  \bibinfo{booktitle}{\emph{Proceedings of the 2012 ACM Conference on Computer
  and Communications Security}}.
\newblock


\bibitem[\protect\citeauthoryear{Katkuri}{Katkuri}{2018}]%
        {katkuri2018survey}
\bibfield{author}{\bibinfo{person}{Sunny Katkuri}.}
  \bibinfo{year}{2018}\natexlab{}.
\newblock \showarticletitle{A survey of data transfer and storage techniques in
  prevalent cryptocurrencies and suggested improvements}.
\newblock \bibinfo{journal}{\emph{arXiv preprint arXiv:1808.03380}}
  (\bibinfo{year}{2018}).
\newblock


\bibitem[\protect\citeauthoryear{Kiayias, Russell, David, and
  Oliynykov}{Kiayias et~al\mbox{.}}{2017}]%
        {kiayias2017ouroboros}
\bibfield{author}{\bibinfo{person}{Aggelos Kiayias}, \bibinfo{person}{Alexander
  Russell}, \bibinfo{person}{Bernardo David}, {and} \bibinfo{person}{Roman
  Oliynykov}.} \bibinfo{year}{2017}\natexlab{}.
\newblock \showarticletitle{Ouroboros: A provably secure proof-of-stake
  blockchain protocol}. In \bibinfo{booktitle}{\emph{Annual International
  Cryptology Conference}}. Springer, \bibinfo{pages}{357--388}.
\newblock


\bibitem[\protect\citeauthoryear{Kim, Ma, Murali, Mason, Miller, and
  Bailey}{Kim et~al\mbox{.}}{2018}]%
        {kim2018measuring}
\bibfield{author}{\bibinfo{person}{Seoung~Kyun Kim}, \bibinfo{person}{Zane Ma},
  \bibinfo{person}{Siddharth Murali}, \bibinfo{person}{Joshua Mason},
  \bibinfo{person}{Andrew Miller}, {and} \bibinfo{person}{Michael Bailey}.}
  \bibinfo{year}{2018}\natexlab{}.
\newblock \showarticletitle{Measuring Ethereum network peers}. In
  \bibinfo{booktitle}{\emph{Proceedings of the Internet Measurement Conference
  2018}}. \bibinfo{pages}{91--104}.
\newblock


\bibitem[\protect\citeauthoryear{Klarman, Basu, Kuzmanovic, and Sirer}{Klarman
  et~al\mbox{.}}{2018}]%
        {bloxroute}
\bibfield{author}{\bibinfo{person}{Uri Klarman}, \bibinfo{person}{Soumya Basu},
  \bibinfo{person}{Aleksandar Kuzmanovic}, {and} \bibinfo{person}{Emin~G{\"u}n
  Sirer}.} \bibinfo{year}{2018}\natexlab{}.
\newblock \showarticletitle{bloxroute: A scalable trustless blockchain
  distribution network whitepaper}.
\newblock \bibinfo{journal}{\emph{IEEE Internet of Things Journal}}
  (\bibinfo{year}{2018}).
\newblock


\bibitem[\protect\citeauthoryear{Kogias, Jovanovic, Gailly, Khoffi, Gasser, and
  Ford}{Kogias et~al\mbox{.}}{2016}]%
        {kogias2016enhancing}
\bibfield{author}{\bibinfo{person}{Eleftherios~Kokoris Kogias},
  \bibinfo{person}{Philipp Jovanovic}, \bibinfo{person}{Nicolas Gailly},
  \bibinfo{person}{Ismail Khoffi}, \bibinfo{person}{Linus Gasser}, {and}
  \bibinfo{person}{Bryan Ford}.} \bibinfo{year}{2016}\natexlab{}.
\newblock \showarticletitle{Enhancing bitcoin security and performance with
  strong consistency via collective signing}. In \bibinfo{booktitle}{\emph{25th
  $\{$USENIX$\}$ Security Symposium ($\{$USENIX$\}$ Security 16)}}.
  \bibinfo{pages}{279--296}.
\newblock


\bibitem[\protect\citeauthoryear{Luu, Narayanan, Zheng, Baweja, Gilbert, and
  Saxena}{Luu et~al\mbox{.}}{2016}]%
        {luu2016secure}
\bibfield{author}{\bibinfo{person}{Loi Luu}, \bibinfo{person}{Viswesh
  Narayanan}, \bibinfo{person}{Chaodong Zheng}, \bibinfo{person}{Kunal Baweja},
  \bibinfo{person}{Seth Gilbert}, {and} \bibinfo{person}{Prateek Saxena}.}
  \bibinfo{year}{2016}\natexlab{}.
\newblock \showarticletitle{A secure sharding protocol for open blockchains}.
  In \bibinfo{booktitle}{\emph{Proceedings of the 2016 ACM SIGSAC Conference on
  Computer and Communications Security}}. \bibinfo{pages}{17--30}.
\newblock


\bibitem[\protect\citeauthoryear{Luu, Velner, Teutsch, and Saxena}{Luu
  et~al\mbox{.}}{2017}]%
        {luu2017smartpool}
\bibfield{author}{\bibinfo{person}{Loi Luu}, \bibinfo{person}{Yaron Velner},
  \bibinfo{person}{Jason Teutsch}, {and} \bibinfo{person}{Prateek Saxena}.}
  \bibinfo{year}{2017}\natexlab{}.
\newblock \showarticletitle{Smartpool: Practical decentralized pooled mining}.
  In \bibinfo{booktitle}{\emph{26th $\{$USENIX$\}$ Security Symposium}}.
\newblock


\bibitem[\protect\citeauthoryear{Malavolta, Moreno-Sanchez, Kate, and
  Maffei}{Malavolta et~al\mbox{.}}{2017}]%
        {silentwhispers}
\bibfield{author}{\bibinfo{person}{Giulio Malavolta}, \bibinfo{person}{Pedro
  Moreno-Sanchez}, \bibinfo{person}{Aniket Kate}, {and} \bibinfo{person}{Matteo
  Maffei}.} \bibinfo{year}{2017}\natexlab{}.
\newblock \showarticletitle{SilentWhispers: Enforcing security and privacy in
  credit networks}. In \bibinfo{booktitle}{\emph{Network and Distributed System
  Security Symposium}}.
\newblock


\bibitem[\protect\citeauthoryear{Malavolta, Moreno-Sanchez, Schneidewind, Kate,
  and Maffei}{Malavolta et~al\mbox{.}}{2019}]%
        {malavolta2019anonymous}
\bibfield{author}{\bibinfo{person}{Giulio Malavolta}, \bibinfo{person}{Pedro
  Moreno-Sanchez}, \bibinfo{person}{Clara Schneidewind},
  \bibinfo{person}{Aniket Kate}, {and} \bibinfo{person}{Matteo Maffei}.}
  \bibinfo{year}{2019}\natexlab{}.
\newblock \showarticletitle{Anonymous Multi-Hop Locks for Blockchain
  Scalability and Interoperability.}. In \bibinfo{booktitle}{\emph{NDSS}}.
\newblock


\bibitem[\protect\citeauthoryear{Maymounkov and Mazieres}{Maymounkov and
  Mazieres}{2002}]%
        {maymounkov2002kademlia}
\bibfield{author}{\bibinfo{person}{Petar Maymounkov} {and}
  \bibinfo{person}{David Mazieres}.} \bibinfo{year}{2002}\natexlab{}.
\newblock \showarticletitle{Kademlia: A peer-to-peer information system based
  on the xor metric}. In \bibinfo{booktitle}{\emph{International Workshop on
  Peer-to-Peer Systems}}. Springer, \bibinfo{pages}{53--65}.
\newblock


\bibitem[\protect\citeauthoryear{McCorry, Bakshi, Bentov, Meiklejohn, and
  Miller}{McCorry et~al\mbox{.}}{2019}]%
        {mccorry2019pisa}
\bibfield{author}{\bibinfo{person}{Patrick McCorry}, \bibinfo{person}{Surya
  Bakshi}, \bibinfo{person}{Iddo Bentov}, \bibinfo{person}{Sarah Meiklejohn},
  {and} \bibinfo{person}{Andrew Miller}.} \bibinfo{year}{2019}\natexlab{}.
\newblock \showarticletitle{Pisa: Arbitration outsourcing for state channels}.
  In \bibinfo{booktitle}{\emph{Proceedings of the 1st ACM Conference on
  Advances in Financial Technologies}}. \bibinfo{pages}{16--30}.
\newblock


\bibitem[\protect\citeauthoryear{Miller, Juels, Shi, Parno, and Katz}{Miller
  et~al\mbox{.}}{2014}]%
        {miller2014permacoin}
\bibfield{author}{\bibinfo{person}{Andrew Miller}, \bibinfo{person}{Ari Juels},
  \bibinfo{person}{Elaine Shi}, \bibinfo{person}{Bryan Parno}, {and}
  \bibinfo{person}{Jonathan Katz}.} \bibinfo{year}{2014}\natexlab{}.
\newblock \showarticletitle{Permacoin: Repurposing bitcoin work for data
  preservation}. In \bibinfo{booktitle}{\emph{2014 IEEE Symposium on Security
  and Privacy}}. IEEE, \bibinfo{pages}{475--490}.
\newblock


\bibitem[\protect\citeauthoryear{Miller, Litton, Pachulski, Gupta, Levin,
  Spring, and Bhattacharjee}{Miller et~al\mbox{.}}{2015}]%
        {miller2015discovering}
\bibfield{author}{\bibinfo{person}{Andrew Miller}, \bibinfo{person}{James
  Litton}, \bibinfo{person}{Andrew Pachulski}, \bibinfo{person}{Neal Gupta},
  \bibinfo{person}{Dave Levin}, \bibinfo{person}{Neil Spring}, {and}
  \bibinfo{person}{Bobby Bhattacharjee}.} \bibinfo{year}{2015}\natexlab{}.
\newblock \showarticletitle{Discovering bitcoin’s public topology and
  influential nodes}.
\newblock  (\bibinfo{year}{2015}).
\newblock


\bibitem[\protect\citeauthoryear{Nayak, Kumar, Miller, and Shi}{Nayak
  et~al\mbox{.}}{2016}]%
        {nayak2016stubborn}
\bibfield{author}{\bibinfo{person}{Kartik Nayak}, \bibinfo{person}{Srijan
  Kumar}, \bibinfo{person}{Andrew Miller}, {and} \bibinfo{person}{Elaine Shi}.}
  \bibinfo{year}{2016}\natexlab{}.
\newblock \showarticletitle{Stubborn mining: Generalizing selfish mining and
  combining with an eclipse attack}. In \bibinfo{booktitle}{\emph{2016 IEEE
  European Symposium on Security and Privacy (EuroS\&P)}}. IEEE,
  \bibinfo{pages}{305--320}.
\newblock


\bibitem[\protect\citeauthoryear{Network}{Network}{2017}]%
        {Bitcoin-relay}
\bibfield{author}{\bibinfo{person}{Bitcoin~Relay Network}.}
  \bibinfo{year}{2017}\natexlab{}.
\newblock \bibinfo{booktitle}{\emph{high-speed block-relay system for miners}}.
\newblock
\urldef\tempurl%
\url{http://www.bitcoinrelaynetwork.org/}
\showURL{%
\tempurl}


\bibitem[\protect\citeauthoryear{Neudecker}{Neudecker}{[n.d.]}]%
        {neudeckercharacterization}
\bibfield{author}{\bibinfo{person}{Till Neudecker}.}
  \bibinfo{year}{[n.d.]}\natexlab{}.
\newblock \showarticletitle{Characterization of the Bitcoin Peer-to-Peer
  Network (2015-2018)}.
\newblock  (\bibinfo{year}{[n.\,d.]}).
\newblock


\bibitem[\protect\citeauthoryear{Neudecker and Hartenstein}{Neudecker and
  Hartenstein}{2017}]%
        {neudecker2017could}
\bibfield{author}{\bibinfo{person}{Till Neudecker} {and}
  \bibinfo{person}{Hannes Hartenstein}.} \bibinfo{year}{2017}\natexlab{}.
\newblock \showarticletitle{Could network information facilitate address
  clustering in Bitcoin?}. In \bibinfo{booktitle}{\emph{International
  conference on financial cryptography and data security}}. Springer,
  \bibinfo{pages}{155--169}.
\newblock


\bibitem[\protect\citeauthoryear{Neudecker and Hartenstein}{Neudecker and
  Hartenstein}{2018}]%
        {neudecker2018network}
\bibfield{author}{\bibinfo{person}{Till Neudecker} {and}
  \bibinfo{person}{Hannes Hartenstein}.} \bibinfo{year}{2018}\natexlab{}.
\newblock \showarticletitle{Network layer aspects of permissionless
  blockchains}.
\newblock \bibinfo{journal}{\emph{IEEE Communications Surveys \& Tutorials}}
  \bibinfo{volume}{21}, \bibinfo{number}{1} (\bibinfo{year}{2018}),
  \bibinfo{pages}{838--857}.
\newblock


\bibitem[\protect\citeauthoryear{Nguyen, Jourjon, Potop-Butucaru, and
  Thai}{Nguyen et~al\mbox{.}}{2019}]%
        {nguyen2019impact}
\bibfield{author}{\bibinfo{person}{Thanh Son~Lam Nguyen},
  \bibinfo{person}{Guillaume Jourjon}, \bibinfo{person}{Maria Potop-Butucaru},
  {and} \bibinfo{person}{Kim Thai}.} \bibinfo{year}{2019}\natexlab{}.
\newblock \showarticletitle{Impact of network delays on Hyperledger Fabric}.
\newblock \bibinfo{journal}{\emph{arXiv preprint arXiv:1903.08856}}
  (\bibinfo{year}{2019}).
\newblock


\bibitem[\protect\citeauthoryear{Otsuki, Aoki, Banno, and Shudo}{Otsuki
  et~al\mbox{.}}{2019}]%
        {otsuki2019effects}
\bibfield{author}{\bibinfo{person}{Kai Otsuki}, \bibinfo{person}{Yusuke Aoki},
  \bibinfo{person}{Ryohei Banno}, {and} \bibinfo{person}{Kazuyuki Shudo}.}
  \bibinfo{year}{2019}\natexlab{}.
\newblock \showarticletitle{Effects of a Simple Relay Network on the Bitcoin
  Network}. In \bibinfo{booktitle}{\emph{Proceedings of the Asian Internet
  Engineering Conference}}. \bibinfo{pages}{41--46}.
\newblock


\bibitem[\protect\citeauthoryear{Ozisik, Andresen, Levine, Tapp, Bissias, and
  Katkuri}{Ozisik et~al\mbox{.}}{2019}]%
        {ozisik2019graphene}
\bibfield{author}{\bibinfo{person}{A~Pinar Ozisik}, \bibinfo{person}{Gavin
  Andresen}, \bibinfo{person}{Brian~N Levine}, \bibinfo{person}{Darren Tapp},
  \bibinfo{person}{George Bissias}, {and} \bibinfo{person}{Sunny Katkuri}.}
  \bibinfo{year}{2019}\natexlab{}.
\newblock \showarticletitle{Graphene: efficient interactive set reconciliation
  applied to blockchain propagation}. In \bibinfo{booktitle}{\emph{Proceedings
  of the ACM Special Interest Group on Data Communication}}.
  \bibinfo{pages}{303--317}.
\newblock


\bibitem[\protect\citeauthoryear{Park, Pietrzak, Kwon, Alwen, Fuchsbauer, and
  Gazi}{Park et~al\mbox{.}}{2018}]%
        {park2018spacemint}
\bibfield{author}{\bibinfo{person}{Sunoo Park}, \bibinfo{person}{Krzysztof
  Pietrzak}, \bibinfo{person}{Albert Kwon}, \bibinfo{person}{Jo{\"e}l Alwen},
  \bibinfo{person}{Georg Fuchsbauer}, {and} \bibinfo{person}{Peter Gazi}.}
  \bibinfo{year}{2018}\natexlab{}.
\newblock \showarticletitle{Spacemint: A cryptocurrency based on proofs of
  space}.
\newblock \bibinfo{journal}{\emph{Financial Cryptography and Data Security}}
  (\bibinfo{year}{2018}).
\newblock


\bibitem[\protect\citeauthoryear{Perkins et~al\mbox{.}}{Perkins
  et~al\mbox{.}}{2001}]%
        {perkins2001ad}
\bibfield{author}{\bibinfo{person}{Charles~E Perkins} {et~al\mbox{.}}}
  \bibinfo{year}{2001}\natexlab{}.
\newblock \bibinfo{booktitle}{\emph{Ad hoc networking}}.
  Vol.~\bibinfo{volume}{1}.
\newblock \bibinfo{publisher}{Addison-wesley Reading}.
\newblock


\bibitem[\protect\citeauthoryear{Pool}{Pool}{2019}]%
        {stratumslushpool}
\bibfield{author}{\bibinfo{person}{Slush Pool}.}
  \bibinfo{year}{2019}\natexlab{}.
\newblock \bibinfo{booktitle}{\emph{Stratum Mining Protocol}}.
\newblock
\urldef\tempurl%
\url{https://slushpool.com/help/stratum-protocol/}
\showURL{%
\tempurl}


\bibitem[\protect\citeauthoryear{Poon and Dryja}{Poon and Dryja}{2016}]%
        {lightning}
\bibfield{author}{\bibinfo{person}{Joseph Poon} {and} \bibinfo{person}{Thaddeus
  Dryja}.} \bibinfo{year}{2016}\natexlab{}.
\newblock \bibinfo{title}{The bitcoin lightning network: Scalable off-chain
  instant payments}.
\newblock
\newblock


\bibitem[\protect\citeauthoryear{Prihodko, Zhigulin, Sahno, Ostrovskiy, and
  Osuntokun}{Prihodko et~al\mbox{.}}{2016}]%
        {prihodko2016flare}
\bibfield{author}{\bibinfo{person}{Pavel Prihodko}, \bibinfo{person}{Slava
  Zhigulin}, \bibinfo{person}{Mykola Sahno}, \bibinfo{person}{Aleksei
  Ostrovskiy}, {and} \bibinfo{person}{Olaoluwa Osuntokun}.}
  \bibinfo{year}{2016}\natexlab{}.
\newblock \showarticletitle{Flare: An approach to routing in lightning
  network}.
\newblock \bibinfo{journal}{\emph{White Paper}} (\bibinfo{year}{2016}).
\newblock


\bibitem[\protect\citeauthoryear{{Raiden Homepage}}{{Raiden Homepage}}{2019}]%
        {raiden}
\bibfield{author}{\bibinfo{person}{{Raiden Homepage}}.}
  \bibinfo{year}{2019}\natexlab{}.
\newblock \bibinfo{title}{The Raiden Network \url{https://raiden.network/}}.
\newblock
\newblock


\bibitem[\protect\citeauthoryear{Recabarren and Carbunar}{Recabarren and
  Carbunar}{2017}]%
        {recabarren2017hardening}
\bibfield{author}{\bibinfo{person}{Ruben Recabarren} {and}
  \bibinfo{person}{Bogdan Carbunar}.} \bibinfo{year}{2017}\natexlab{}.
\newblock \showarticletitle{Hardening stratum, the bitcoin pool mining
  protocol}.
\newblock \bibinfo{journal}{\emph{Proceedings on Privacy Enhancing
  Technologies}} \bibinfo{volume}{2017}, \bibinfo{number}{3}
  (\bibinfo{year}{2017}), \bibinfo{pages}{57--74}.
\newblock


\bibitem[\protect\citeauthoryear{Ripple}{Ripple}{2019}]%
        {ripple-overlay}
\bibfield{author}{\bibinfo{person}{Ripple}.} \bibinfo{year}{2019}\natexlab{}.
\newblock \bibinfo{booktitle}{\emph{Overlay}}.
\newblock
\urldef\tempurl%
\url{https://github.com/ripple/rippled/tree/develop/src/ripple/overlay}
\showURL{%
\tempurl}


\bibitem[\protect\citeauthoryear{Rivest}{Rivest}{1997}]%
        {rivest1997electronic}
\bibfield{author}{\bibinfo{person}{Ronald~L Rivest}.}
  \bibinfo{year}{1997}\natexlab{}.
\newblock \showarticletitle{Electronic lottery tickets as micropayments}. In
  \bibinfo{booktitle}{\emph{International Conference on Financial
  Cryptography}}. Springer, \bibinfo{pages}{307--314}.
\newblock


\bibitem[\protect\citeauthoryear{Rohrer, Malliaris, and Tschorsch}{Rohrer
  et~al\mbox{.}}{2019}]%
        {rohrer2019discharged}
\bibfield{author}{\bibinfo{person}{Elias Rohrer}, \bibinfo{person}{Julian
  Malliaris}, {and} \bibinfo{person}{Florian Tschorsch}.}
  \bibinfo{year}{2019}\natexlab{}.
\newblock \showarticletitle{Discharged Payment Channels: Quantifying the
  Lightning Network's Resilience to Topology-Based Attacks}.
\newblock \bibinfo{journal}{\emph{arXiv preprint arXiv:1904.10253}}
  (\bibinfo{year}{2019}).
\newblock


\bibitem[\protect\citeauthoryear{Roos, Moreno-Sanchez, Kate, and Goldberg}{Roos
  et~al\mbox{.}}{2017}]%
        {roos2017settling}
\bibfield{author}{\bibinfo{person}{Stefanie Roos}, \bibinfo{person}{Pedro
  Moreno-Sanchez}, \bibinfo{person}{Aniket Kate}, {and} \bibinfo{person}{Ian
  Goldberg}.} \bibinfo{year}{2017}\natexlab{}.
\newblock \showarticletitle{Settling payments fast and private: Efficient
  decentralized routing for path-based transactions}.
\newblock \bibinfo{journal}{\emph{arXiv preprint arXiv:1709.05748}}
  (\bibinfo{year}{2017}).
\newblock


\bibitem[\protect\citeauthoryear{Sapirshtein, Sompolinsky, and
  Zohar}{Sapirshtein et~al\mbox{.}}{2016}]%
        {sapirshtein2016optimal}
\bibfield{author}{\bibinfo{person}{Ayelet Sapirshtein},
  \bibinfo{person}{Yonatan Sompolinsky}, {and} \bibinfo{person}{Aviv Zohar}.}
  \bibinfo{year}{2016}\natexlab{}.
\newblock \showarticletitle{Optimal selfish mining strategies in bitcoin}. In
  \bibinfo{booktitle}{\emph{International Conference on Financial Cryptography
  and Data Security}}.
\newblock


\bibitem[\protect\citeauthoryear{Sivaraman, Venkatakrishnan, Alizadeh, Fanti,
  and Viswanath}{Sivaraman et~al\mbox{.}}{2018}]%
        {sivaraman2018routing}
\bibfield{author}{\bibinfo{person}{Vibhaalakshmi Sivaraman},
  \bibinfo{person}{Shaileshh~Bojja Venkatakrishnan}, \bibinfo{person}{Mohammad
  Alizadeh}, \bibinfo{person}{Giulia Fanti}, {and} \bibinfo{person}{Pramod
  Viswanath}.} \bibinfo{year}{2018}\natexlab{}.
\newblock \showarticletitle{Routing cryptocurrency with the spider network}.
\newblock \bibinfo{journal}{\emph{arXiv preprint arXiv:1809.05088}}
  (\bibinfo{year}{2018}).
\newblock


\bibitem[\protect\citeauthoryear{Tang, Wang, Fanti, and Oh}{Tang
  et~al\mbox{.}}{2019}]%
        {tang2019privacy}
\bibfield{author}{\bibinfo{person}{Weizhao Tang}, \bibinfo{person}{Weina Wang},
  \bibinfo{person}{Giulia Fanti}, {and} \bibinfo{person}{Sewoong Oh}.}
  \bibinfo{year}{2019}\natexlab{}.
\newblock \showarticletitle{Privacy-Utility Tradeoffs in Routing Cryptocurrency
  over Payment Channel Networks}.
\newblock \bibinfo{journal}{\emph{arXiv preprint arXiv:1909.02717}}
  (\bibinfo{year}{2019}).
\newblock


\bibitem[\protect\citeauthoryear{Tochner and Schmid}{Tochner and
  Schmid}{2020}]%
        {saar-economy}
\bibfield{author}{\bibinfo{person}{Saar Tochner} {and} \bibinfo{person}{Stefan
  Schmid}.} \bibinfo{year}{2020}\natexlab{}.
\newblock \showarticletitle{On Search Friction of Route Discovery in Offchain
  Networks}.
\newblock \bibinfo{journal}{\emph{arXiv preprint arXiv:2005.14676}}
  (\bibinfo{year}{2020}).
\newblock


\bibitem[\protect\citeauthoryear{Tochner, Schmid, and Zohar}{Tochner
  et~al\mbox{.}}{2019}]%
        {tochner2019hijacking}
\bibfield{author}{\bibinfo{person}{Saar Tochner}, \bibinfo{person}{Stefan
  Schmid}, {and} \bibinfo{person}{Aviv Zohar}.}
  \bibinfo{year}{2019}\natexlab{}.
\newblock \showarticletitle{Hijacking Routes in Payment Channel Networks: A
  Predictability Tradeoff}.
\newblock \bibinfo{journal}{\emph{arXiv preprint arXiv:1909.06890}}
  (\bibinfo{year}{2019}).
\newblock


\bibitem[\protect\citeauthoryear{Tochner and Zohar}{Tochner and Zohar}{2018}]%
        {tochner2018pick}
\bibfield{author}{\bibinfo{person}{Saar Tochner} {and} \bibinfo{person}{Aviv
  Zohar}.} \bibinfo{year}{2018}\natexlab{}.
\newblock \showarticletitle{How to pick your friends-a game theoretic approach
  to p2p overlay construction}.
\newblock \bibinfo{journal}{\emph{arXiv preprint arXiv:1810.05447}}
  (\bibinfo{year}{2018}).
\newblock


\bibitem[\protect\citeauthoryear{Tran, Choi, Moon, Vu, and Kang}{Tran
  et~al\mbox{.}}{2020}]%
        {tran2020stealthier}
\bibfield{author}{\bibinfo{person}{Muoi Tran}, \bibinfo{person}{Inho Choi},
  \bibinfo{person}{Gi~Jun Moon}, \bibinfo{person}{Anh~V Vu}, {and}
  \bibinfo{person}{Min~Suk Kang}.} \bibinfo{year}{2020}\natexlab{}.
\newblock \showarticletitle{A Stealthier Partitioning Attack against Bitcoin
  Peer-to-Peer Network}.
\newblock  (\bibinfo{year}{2020}).
\newblock


\bibitem[\protect\citeauthoryear{Troncoso, Isaakidis, Danezis, and
  Halpin}{Troncoso et~al\mbox{.}}{2017}]%
        {troncoso2017systematizing}
\bibfield{author}{\bibinfo{person}{Carmela Troncoso}, \bibinfo{person}{Marios
  Isaakidis}, \bibinfo{person}{George Danezis}, {and} \bibinfo{person}{Harry
  Halpin}.} \bibinfo{year}{2017}\natexlab{}.
\newblock \showarticletitle{Systematizing decentralization and privacy: Lessons
  from 15 years of research and deployments}.
\newblock \bibinfo{journal}{\emph{Proceedings on Privacy Enhancing
  Technologies}} \bibinfo{volume}{2017}, \bibinfo{number}{4}
  (\bibinfo{year}{2017}), \bibinfo{pages}{404--426}.
\newblock


\bibitem[\protect\citeauthoryear{Tschipper}{Tschipper}{2016}]%
        {tschipper2016buip010}
\bibfield{author}{\bibinfo{person}{Peter Tschipper}.}
  \bibinfo{year}{2016}\natexlab{}.
\newblock \showarticletitle{BUIP010: Xtreme Thinblocks}. In
  \bibinfo{booktitle}{\emph{Bitcoin Forum (1 January 2016). https://bitco.
  in/forum/threads/buip010-passed-xtreme-thinblocks}},
  Vol.~\bibinfo{volume}{774}.
\newblock


\bibitem[\protect\citeauthoryear{Utz~Nisslmueller and Decker}{Utz~Nisslmueller
  and Decker}{2020}]%
        {icissp20}
\bibfield{author}{\bibinfo{person}{Stefan~Schmid Utz~Nisslmueller,
  Klaus-Tycho~Foerster} {and} \bibinfo{person}{Christian Decker}.}
  \bibinfo{year}{2020}\natexlab{}.
\newblock \showarticletitle{Toward Active and Passive Confidentiality Attacks
  On Cryptocurrency Off-Chain Networks}. In \bibinfo{booktitle}{\emph{Proc. 6th
  International Conference on Information Systems Security and Privacy
  (ICISSP)}}.
\newblock


\bibitem[\protect\citeauthoryear{Vasek, Thornton, and Moore}{Vasek
  et~al\mbox{.}}{2014}]%
        {vasek2014empirical}
\bibfield{author}{\bibinfo{person}{Marie Vasek}, \bibinfo{person}{Micah
  Thornton}, {and} \bibinfo{person}{Tyler Moore}.}
  \bibinfo{year}{2014}\natexlab{}.
\newblock \showarticletitle{Empirical analysis of denial-of-service attacks in
  the Bitcoin ecosystem}. In \bibinfo{booktitle}{\emph{International conference
  on financial cryptography and data security}}. Springer,
  \bibinfo{pages}{57--71}.
\newblock


\bibitem[\protect\citeauthoryear{Vorkapic}{Vorkapic}{2018}]%
        {vorkapic2018secure}
\bibfield{author}{\bibinfo{person}{Aleksandar Vorkapic}.}
  \bibinfo{year}{2018}\natexlab{}.
\newblock \bibinfo{title}{Secure Blockchain Network Communication using SCION}.
\newblock
\newblock


\bibitem[\protect\citeauthoryear{Wang, Chu, and Yang}{Wang
  et~al\mbox{.}}{2019}]%
        {wang2019measurement}
\bibfield{author}{\bibinfo{person}{Canhui Wang}, \bibinfo{person}{Xiaowen Chu},
  {and} \bibinfo{person}{Qin Yang}.} \bibinfo{year}{2019}\natexlab{}.
\newblock \showarticletitle{Measurement and analysis of the bitcoin networks: A
  view from mining pools}.
\newblock \bibinfo{journal}{\emph{arXiv preprint arXiv:1902.07549}}
  (\bibinfo{year}{2019}).
\newblock


\bibitem[\protect\citeauthoryear{W{\"u}st and Gervais}{W{\"u}st and
  Gervais}{2016}]%
        {wust2016Ethereum}
\bibfield{author}{\bibinfo{person}{Karl W{\"u}st} {and} \bibinfo{person}{Arthur
  Gervais}.} \bibinfo{year}{2016}\natexlab{}.
\newblock \bibinfo{booktitle}{\emph{Ethereum eclipse attacks}}.
\newblock \bibinfo{type}{{T}echnical {R}eport}. \bibinfo{institution}{ETH
  Zurich}.
\newblock


\bibitem[\protect\citeauthoryear{Yang and Garcia-Molina}{Yang and
  Garcia-Molina}{2003}]%
        {yang2003ppay}
\bibfield{author}{\bibinfo{person}{Beverly Yang} {and} \bibinfo{person}{Hector
  Garcia-Molina}.} \bibinfo{year}{2003}\natexlab{}.
\newblock \showarticletitle{PPay: micropayments for peer-to-peer systems}. In
  \bibinfo{booktitle}{\emph{Proceedings of the 10th ACM conference on Computer
  and communications security}}. \bibinfo{pages}{300--310}.
\newblock


\bibitem[\protect\citeauthoryear{Young, Kate, Goldberg, and Karsten}{Young
  et~al\mbox{.}}{2013}]%
        {young2013towards}
\bibfield{author}{\bibinfo{person}{Maxwell Young}, \bibinfo{person}{Aniket
  Kate}, \bibinfo{person}{Ian Goldberg}, {and} \bibinfo{person}{Martin
  Karsten}.} \bibinfo{year}{2013}\natexlab{}.
\newblock \showarticletitle{Towards practical communication in
  Byzantine-resistant DHTs}.
\newblock \bibinfo{journal}{\emph{IEEE/ACM Transactions on Networking (ToN)}}
  \bibinfo{volume}{21}, \bibinfo{number}{1} (\bibinfo{year}{2013}),
  \bibinfo{pages}{190--203}.
\newblock


\bibitem[\protect\citeauthoryear{Zamyatin, Stifter, Schindler, Weippl, and
  Knottenbelt}{Zamyatin et~al\mbox{.}}{2018}]%
        {zamyatin2018flux}
\bibfield{author}{\bibinfo{person}{Alexei Zamyatin}, \bibinfo{person}{Nicholas
  Stifter}, \bibinfo{person}{Philipp Schindler}, \bibinfo{person}{Edgar~R
  Weippl}, {and} \bibinfo{person}{William~J Knottenbelt}.}
  \bibinfo{year}{2018}\natexlab{}.
\newblock \showarticletitle{Flux: Revisiting Near Blocks for Proof-of-Work
  Blockchains.}
\newblock \bibinfo{journal}{\emph{IACR Cryptology ePrint Archive}}
  \bibinfo{volume}{2018} (\bibinfo{year}{2018}), \bibinfo{pages}{415}.
\newblock


\end{thebibliography}
}

\end{document}